\documentclass[10pt,twocolumn]{article}

\usepackage{fullpage}

\usepackage{graphicx}
\usepackage{multirow}
\usepackage{array}
\usepackage{flushend}
\usepackage{fancyvrb}
\usepackage{times}
\usepackage{amsmath}
\usepackage{amsfonts}
\usepackage{algorithm}
\usepackage{appendix}
\usepackage[noend]{algpseudocode}

\usepackage[pdftex, colorlinks=true, citecolor=black, linkcolor=black,
citebordercolor={0 0 0}, urlcolor=black]{hyperref}

\newcommand{\mycomment}[1]{}
\newcommand{\ignore}[1]{}
\ifpdf
\renewcommand{\url}[1]{\texttt{\small{#1}}}
\else
\newcommand{\url}[1]{\texttt{\small{#1}}}
\fi
\newcommand{\code}[1]{\texttt{#1}}
\newcommand{\codesm}[1]{\texttt{\small #1}}

\newcommand{\toolshort}{ASAP}

\newcommand{\reducename}{stochastic reduce}

\makeatletter
\def\BState{\State\hskip-\ALG@thistlm}
\makeatother

\usepackage{bbm, bm}

\numberwithin{equation}{section}

\newcommand{\mb}{\mathbf}
\renewcommand{\mb}{\boldsymbol}
\renewcommand{\bm}{\boldsymbol}
\newcommand{\mc}{\mathcal}

\newcommand{\norm}[2]{\left\| #1 \right\|_{#2}}
\newcommand{\reals}{\mathbb{R}}

\newcommand{\set}[1]{\left\{ #1 \right\}}

\newlength{\dhatheight}

\newcommand{\diag}[1]{\mathrm{diag}\left( #1 \right)}

\begin{document}

\title{\bf ASAP: Asynchronous Approximate Data-Parallel Computation}
\author{Asim Kadav and Erik Kruus\\NEC Labs, Princeton \\}
\date{}
\maketitle
\thispagestyle{empty}

\begin{abstract}
Emerging workloads, such as graph processing and machine learning are approximate because of the scale of data involved and the stochastic nature of the underlying algorithms. These algorithms are often distributed over
multiple machines using bulk-synchronous processing (BSP) or other synchronous processing paradigms such as map-reduce. However, data parallel processing primitives such as repeated \codesm{barrier} and \codesm{reduce} operations introduce high synchronization overheads. Hence, many existing data-processing platforms use asynchrony and staleness to improve data-parallel job performance. Often, these systems simply change the synchronous communication to asynchronous between the worker nodes in the cluster. This improves the throughput of data processing but results in poor accuracy of the final output since different workers may progress at different speeds and process inconsistent intermediate outputs.

In this paper, we present \toolshort, a model that provides asynchronous and approximate processing semantics for data-parallel computation. \toolshort\ provides fine-grained worker synchronization 
using {\codesm {NOTIFY-ACK}} semantics that allows independent workers to run asynchronously. \toolshort\ also provides {\em \reducename} that provides approximate but guaranteed convergence to the same result as an aggregated all-reduce. In our results, we show that \toolshort\ can reduce synchronization costs and provides 2-10X speedups in convergence and up to 10X savings in network costs for distributed machine learning applications and provides strong convergence guarantees.


\end{abstract}


\section{Introduction}
\label{sec:intro}
Large-scale data-parallel computation across multiple machines often provide two fundamental constructs to scale-out local data processing. First, a {\it merge} or a {\it reduce} operation to allow the workers to merge updates from all other workers and second, a {\it barrier} or an implicit {\it wait}  operation to ensure that all workers can synchronize and operate at similar speeds. For example, the bulk-synchronous model is a general paradigm to model data-intensive distributed processing~\cite{valiant1990bridging}. Here, each node after processing a specific amount of data synchronizes with the other nodes using \codesm{barrier} and \codesm{reduce} operations. The BSP model is widely used to implement many big data applications, frameworks and libraries such as in the areas of graph processing and machine learning~\cite{apachehama:web, gabriel2004open, gonzalez2012powergraph, gregor2005parallel, malewicz2010pregel,power2010piccolo}. Other synchronous paradigms such as the map-reduce~\cite{dean2008mapreduce, zaharia2012resilient}, the parameter server~\cite{dean2012large, li2014parameterserver} and dataflow based systems~\cite{abaditensorflow, isard2007dryad, murray2013naiad} use similar constructs to synchronize outputs across multiple workers.

There is an emerging class of big data applications such as graph-processing and machine learning that are approximate because of the stochastic nature of the underlying algorithms that converge to a final solution in an iterative fashion. These iterative-convergent algorithms operate on large amounts of data and unlike traditional TPC style workloads that are CPU bound~\cite{ousterhout2015making}, these iterative algorithms incur significant network and synchronization costs by communicating large vectors between their workers. These applications can gain an increase in performance by reducing the synchronization costs in two ways. First, the workers can operate over stale intermediate outputs. The stochastic algorithms operate over input data in an iterative fashion to produce and communicate intermediate outputs with other workers. However, it is not imperative that the workers 
perform a \codesm{reduce} on all the intermediate outputs at every iteration. Second, the synchronization requirements between the workers may be relaxed, allowing partial, stale or overwritten outputs. This is possible because in some cases the iterative nature of data processing and the stochastic nature of the algorithms may provide an opportunity to correct any errors introduced from staleness or incorrect synchronization.

There has been recent research that explores this opportunity by processing stale outputs, or by removing all synchronization~\cite{recht2011hogwild, wang2013asynchronous}. However, na\"{\i}vely converting the algorithms from synchronous to asynchronous can increase the throughput but may not improve the convergence speeds. 
Furthermore, these bounds may not generalize across applications or datasets. Finally, an increase in the data processing throughput may not translate to an improved speedup with the same final output accuracy and in some cases may even converge the underlying algorithm to an incorrect value~\cite{frank:graphprocessing}.

Hence, to provide asynchronous and approximate semantics with reasonable correctness guarantees for iterative convergent algorithms, we present Asynchronous and Approximate abstractions for data-parallel computation. To facilitate approximate processing, we describe {\em \reducename}, a sparse reduce primitive, that mitigates the communication and synchronization costs by performing the \codesm{reduce} operation with fewer workers. We construct a reduce operator by choosing workers based on sparse {\em expander} graphs on underlying communication nodes that mitigates CPU and network costs during \codesm{reduce} for iterative convergent algorithms. Furthermore, \reducename\ allows developers to reason about convergence times and network communication costs, by evaluating the specific properties of the underlying node communication graphs. 

To reduce synchronization costs, we propose a fine-grained communication using an OS style \codesm{NOTIFY-ACK} mechanism that provides performance improvement over \codesm{barrier} style synchronization. \codesm{NOTIFY-ACK} allows independent worker threads (such as those in \reducename) to run asynchronously instead of blocking on a coarse-grained global \codesm{barrier} at every iteration. Additionally, \codesm{NOTIFY-ACK} provides stronger consistency than just using a \codesm{barrier} to implement synchronous data-parallel processing.

\toolshort\ is not a programming model (like map-reduce~\cite{dean2008mapreduce}) or is limited to  a set of useful implementation mechanisms. It introduces semantics for approximate and asynchronous execution which are amiss in the current flurry of distributed machine learning systems which often use asynchrony and staleness to improve the input processing throughput.

The contributions of this paper are as follows:
\begin{itemize}
\item We present ASAP, an asynchronous approximate computation model for large scale data parallel applications. 
We introduce \reducename\ for approximate semantics and fine-grained synchronization based on \codesm{NOTIFY-ACK} to allow independent threads run asynchronously.

\item We apply ASAP to distributed machine learning. We empirically and formally show that data parallel learning using ASAP semantics converges to the same result as synchronous all-reduce.

\end{itemize}

In our results, we show that when ASAP semantics are applied to distributed machine learning problems, the resulting system can achieve strong consistency, provable convergence and provides 2-10X in convergence and up to 10X savings in network costs. The rest of the paper is organized as follows. In the next section, we provide a background on data-parallel distributed machine learning. We then describe the \toolshort\ model which consists of the \reducename\ and fine-grained synchronization primitives that reduces synchronization costs. Finally, we evaluate \toolshort\ and discuss related work.

\section{Background}
\label{sec:back}
Large-scale problems such as training image classification models, page-rank computation and matrix factorization operate on large amounts of data. As a result, many stochastic algorithms have been proposed that make these problem tractable for large data by iteratively approximating the solution over small batches of data. For example, to scale better, matrix factorization methods have moved from direct and exact factorization methods such as singular value decomposition to iterative and approximate factorization using gradient descent style algorithms~\cite{gemulla2011large}. Hence, many algorithms that are used to discover relationships amongst data have been re-written in the form distributed optimization problems that iterate over the input data and approximate the solution. In this paper, we describe how to provide 
asynchronous and approximate semantics to these distributed optimization based applications. We specifically focus on the design and performance benefits for distributed machine learning applications using \toolshort.



\subsection{Distributed Machine Learning}

\begin{figure}
\centering
\includegraphics[keepaspectratio=true,width=3.2in]{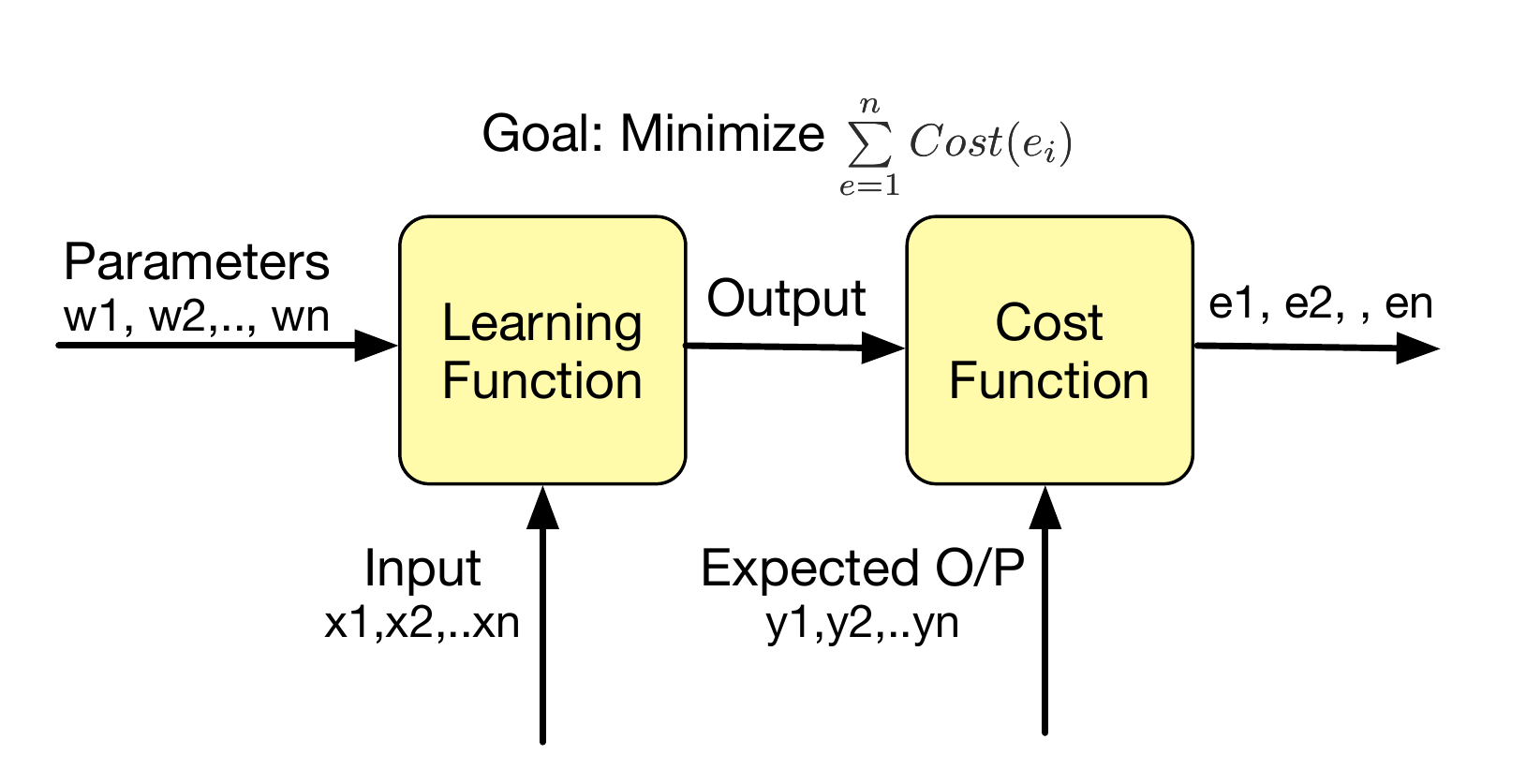}
\caption{\small \bf The machine learning training process. The computation can be distributed over a cluster of machines with data parallelism, by splitting the input ($x1$,$x2$,..,$xn$) or model parallelism, by splitting the model ($w1$, $w2$,..,$wn$). The goal of parallelization is not only to process the input quickly but also to maintain low error rates ($e1$, $e2$,..,$en$).}
\label{fig:ml-problems}
\vspace{-0.2in}
\end{figure}

Machine learning algorithms process data to build a training model that can
generalize over new data. The training output model, or the parameter vector (represented by $w$) is computed to perform future predictions over new data. To train over large data, ML methods often use the Stochastic Gradient Descent (SGD) algorithm that can train over a single (or a batch) of examples over time. The SGD algorithm processes data examples to compute the gradient of a loss function. The parameter vector is then updated based on this gradient value after processing each training data example. After a number of iterations, the parameter vector or the model converges towards acceptable error values over test data. Hence, SGD can be used to train ML models iteratively over the training data as shown in Figure~\ref{fig:ml-problems}.

To scale out the computation over multiple machines, the SGD algorithm can be distributed over a cluster by using data parallelism, by splitting the input ($x1$,$x2$,..,$xn$) or by model parallelism, by splitting the model ($w1$, $w2$,..,$wn$). In data-parallel learning using BSP, the parallel model replicas train over different machines. After a fixed number of 
iterations, these machines synchronize the parameter models that have been trained over the partitioned data with one-another using a \codesm{reduce} operation. For example, each machine may perform an average of all incoming models with its own model, and proceed to train over more data. In the BSP model, there is a global barrier that ensures that models train and synchronize intermediate inputs at the same speeds. 


Hence, distributed data-parallel machine learning suffers from additional synchronization and communication costs over a single thread. The \codesm{reduce} operation requires communicating models to all other machines
when training using the BSP or map-reduce model. However, since these algorithms are {\em iterative-convergent}, and can tolerate errors in the
synchronization step, there has been recent work on communicating stale intermediate parameter updates and exchanging parameters with little or no synchronization~\cite{canny:butterfly, cipar2013solving, recht2011hogwild}.

Past research has found that simply removing the barrier may speed up the throughput of the system~\cite{chilimbi2014adam, wang2013asynchronous}. However, this may not always improve the convergence speed and may even converge the system to an incorrect final value~\cite{frank:graphprocessing}. Since the workers do not synchronize, and communicate model parameters at different speeds, the workers process the data examples rapidly. However, since different workers train at different speeds, the global model may skew in the favor of the workers that are able to process and communicate their models. Similarly, setting bounds for synchrony may appear to work for some workloads. But determining these bounds can be empirically difficult and for some datasets it may be no better than an single iteration (i.e. no better than having a barrier). Furthermore, if a global single model is maintained and updated without locks (as in ~\cite{recht2011hogwild}), a global convergence may only be possible if the parameter vector is sparse. Finally, maintaining a global single model in a distributed setting results in lots of wasted communication since a lot of the useful parameter updates are overwritten~\cite{chilimbi2014adam, noel2014dogwild}.

The distributed parameter-server architecture limits network traffic by
maintaining a central master~\cite{abaditensorflow, dean2012large, li2014parameterserver}. Here, the server coordinates the parameter consistency amongst all other worker machines by resetting the workers' model after every iteration and ensures global consensus on the final model. Hence, a single server communicates with a large number of workers that may result in
network congestion at the edges which can be mitigated using a distributed parameter server~\cite{li2014parameterserver}. However, the parameter server suffers from similar synchronization issues as BSP style systems -- a synchronous server may spend a significant amount of time at the \codesm{barrier} while an asynchronous server may \codesm{reduce} with few workers' models and produce inconsistent intermediate outputs and this can slow down convergence. Hence, the parameter server architecture can benefit from a fine-grained consistent synchronization mechanisms that have low overheads.

To provide asynchronous and approximate processing semantics with consistency and convergence guarantees, we introduce \toolshort\ that provides approximate processing by synchronizing each worker with a subset of workers at each iteration. Additionally, \toolshort\ provides fine-grained synchronization that improves convergence behavior and reduces synchronization overheads over a \codesm{barrier}. We describes both these techniques next.

\section{Stochastic reduce for approximate processing}
\label{sec:partial-reduce}

In this section, we describe how data-parallel applications can use \reducename\ to
mitigate network and processing times for iterative machine learning algorithms. We then introduce a metric
to compare convergence speeds of \reducename\ with \codesm{all-reduce}.

With distributed machine learning, parallel machines or cores (workers) train on model replicas
in parallel and exchange model parameters after processing a set of examples. Hence,
after processing a batch of examples, the parallel model replicas perform a \codesm{reduce}
over the incoming parameters, update their local models and continue to train. In the 
map-reduce model, all workers synchronize with one another and this operation is referred
to as the {\em all-reduce} step. In the parameter server model, this synchronization and \codesm{reduce} occurs with 
a single or distributed master~\cite{dean2012large, li2013distributed}. To mitigate the reduce overheads,
efficient all-reduce has been explored in the map-reduce context where nodes perform partial aggregation 
in a tree-style to reduce network costs~\cite{chu2003optimizing, chaiken2008scope, yu2008dryadlinq}. However, these methods decrease the network and processing costs
at the cost of increasing the latency of the \codesm{reduce} operation proportional to the height of the tree. 

\begin{figure}
\centering
\includegraphics[keepaspectratio=true,width=3.2in]{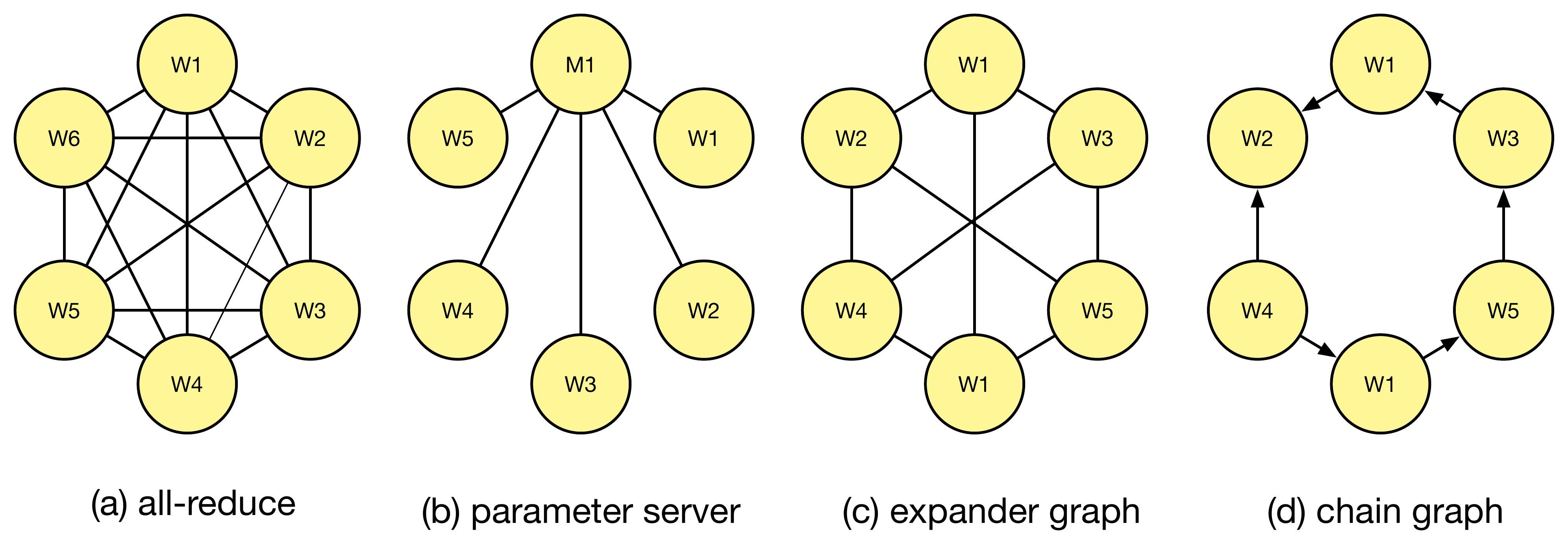}
\caption{\small \bf Figure (a) shows all-reduce, Spectral Gap for six nodes, SG-6:1.00, SG-25:1.00. Figure (b) shows parameter server, SG-6:0.75, SG-25:0.68. Figure (c) shows an expander graph with SG-6:0.38, SG-25:0.2 Figure (d) shows a chain graph with SG-6: 0.1, SG-25: 0.002}
\label{fig:nodegraphs}
\vspace{-0.2in}
\end{figure}

The network of worker nodes, i.e. the node communication graph of the cluster determines how rapidly
the intermediate model parameters are propagated to all other machines and also determines the associated network
and processing costs. For example, all-reduce and the 
parameter server represent different types of communication graphs that describe how the workers 
communicate the intermediate results as shown in Figure \ref{fig:nodegraphs}. 
Intuitively, when the workers communicate with all machines at every \codesm{reduce} step, 
this network is densely connected and convergence is rapid since all the machines get the latest intermediate
updates. However, if the network of nodes is sparsely connected, the convergence may
be slow due to stale, indirect updates being exchanged between machines. However, with sparse connectivity, there are savings in 
network and CPU costs (fewer updates to process at each node), that can result in an overall speedup in job completion times.
Furthermore, if there is a heterogeneity in communication bandwidths between the workers
(or between the workers and the master, if a master is used), many workers may end up 
waiting. As an example, if one is training using GPUs over a cluster, GPUs within
one machine can synchronize at far lower costs over the PCI bus than over the network. Hence, frequent \codesm{reduce}
across interconnects with varying latency may introduce a large number of stragglers which can increase 
the cost of synchronization for all workers. 

Hence, instead of synchronizing and performing a reduce with every other node, parallel nodes
training ML tasks can synchronize with fewer nodes by performing a sparse or {\em \reducename} using
a sparse node communication graph. Since machine learning algorithms that use SGD are stochastic, as long as all 
machines are connected and they receive parameter updates directly or indirectly from others,
the underlying optimization algorithm will converge to the same result as an all-reduce. When the nodes have strong 
connectivity properties, i.e. the nodes are connected such that the parameter updates from one node disperse 
to all other nodes in fewest time steps at low network costs, the convergence is network efficient. 
We propose using \reducename\ based on sparse graphs with strong connectivity properties. 

\subsection{Rapid convergence at fixed communication costs}
The goal of \reducename\ is to improve performance by reducing network and processing costs
by using sparse reduce graphs. Recent work has shown that every dense graph can be 
reduced to an equivalent approximate sparse graph with fewer edges~\cite{batson2013spectral}. This is a significant result since it implies that \reducename\ can be applied to save network costs for almost any network topology. Expander graphs, which are sparse graphs with strong connectivity properties have been explored in the context
of data centers and distributed communities to communicate data with low overheads~\cite{tran2009sybil, valadarsky2015xpander}. An expander graph has fixed out-degrees as the number of vertices increase while maintaining approximately the same connectivity between the vertices. Hence, using expander graphs for \reducename\ provides approximately the same convergence as all-reduce while keeping network costs low, as the number of nodes increase.

To measure the convergence of algorithms that use \reducename, i.e. to compare the sparsity of the adjacency graph of communication, we calculate the {\em spectral gap} of the adjacency matrix of the network of workers. The spectral gap is the difference between the two largest singular values of the adjacency matrix normalized by the in-degree of every node. The spectral gap of a communication graph determines how rapidly a distributed, iterative sparse reduce converges when performing distributed optimization over a network of nodes represented by this graph. For strong convergence properties, this value should be as high as possible. Hence, well-connected graphs have a high spectral gap value and converge faster but will have high communication costs. Conversely, if the graph is disconnected, the spectral gap value is {\em zero}, and with data partitioned across machines, the final model may not converge to a correct value. We discuss the formal conditions of convergence based on spectral gap for any optimization problem using stochastic reduce in Appendix A.

Past work using partial-reduce for optimization problems has explored specific fixed communication graphs such as butterfly or logarithmic sequences~\cite{canny:butterfly, li2015malt}. These communication sequences provide fixed network costs but may not generalize to networks with complex topologies or networks with dissimilar bandwidth edges such as distributed GPU environments. Additionally, \toolshort\ introduces the ability to reason convergence using the spectral gap of a network graph and developers can reason why some node graphs have stronger convergence properties than others. Finally, existing approaches use a global barrier after each communication step and require all machines to wait for all other nodes after each reduce, incurring extra synchronization overheads. We describe how fine-grained communication of \toolshort\ improves reduce this synchronization overhead in Section \ref{sec:async-notify-ack}. We now describe how a developer can generate sparse graphs with good convergence properties given fixed network costs or fixed out-degrees for vertices in the communication graph. 

Figure~\ref{fig:nodegraphs} shows six nodes connected using four distributed machine learning training architectures, the all-reduce, the parameter server (non-distributed), an expander graph with a fixed out-degree of two, and a chain like architecture and their respective spectral-gap values for $6$ and $25$ nodes. As expected, architectures with more edges, have a high-spectral gap and provide good convergence. Figure~\ref{fig:nodegraphs} (a) shows the all-reduce, where all machines communicate with one-another and increasing the number of nodes, significantly increases the network costs. Figure~\ref{fig:nodegraphs} (b) shows the parameter server has a reasonably high spectral gap but using a single master with a high fanout requires considerable network bandwidth and Paxos-style reliability for the master. Figure~\ref{fig:nodegraphs} (c) shows a root expander graph has a fixed out-degree of two, and in a network of $N$ total nodes, each node $i$ sends the intermediate output to its neighbor ($i+1$) (to ensure connectivity) and to $i + \sqrt{N}$th node. Such root expander graphs ensure that the updates are spread across the network as $N$ scales since the root increases with $N$. Finally, figure~\ref{fig:nodegraphs}(d), shows a chain like graph, where the nodes are connected in a chain-like fashion, the intermediate parameter updates from node $i$ may spread to $i+1$ in a single time step, but will require $N$ time steps to reach to the last node in the cluster and has low spectral gap values.
In the rest of the paper, we use the {\em root} sparse expander graph, as shown in Figure~\ref{fig:nodegraphs}(c), with a fixed out-degree of two, to evaluate \reducename. A full discussion of various expander graph architectures and their corresponding convergence is interesting but outside the scope of this paper.

We find that by using sparse expander reduce graphs, with just two out-degrees often provides good convergence and speedup over all-reduce i.e. high enough spectral gap values with reasonably low communication costs. Using sparse directed reduce graphs with \reducename\ results in faster model training times because: First, the amount of network time is reduced. Second, the synchronization time is reduced since each machine communicates with fewer parallel models. Finally, the CPU times at each model replica is reduced since it needs to process fewer incoming intermediate parameter updates.

For \reducename\ to be effective, the following properties are desirable.
First, the generated node communication graph should have a 
{\bf high-spectral gap}. This ensures that the model updates from each machine are diffused across
the network rapidly. Second, the node-communication graphs should have
{\bf low communication costs}. For example, the out-degrees of each node in the
graph should have small out-degrees. Finally, the graph should be   
{\bf easy to generate} such as using a sequence to accommodate variable number 
of machines or a possible re-configuration in case of a failure. These properties
can be used to guide existing data-parallel optimizers~\cite{guo2012spotting} or schedulers~\cite{chowdhury2014efficient},
to reduce data shuffling costs by constructing sparse reduce graphs while accommodating
real world constraints such as avoiding cross-rack or cross-interconnect \codesm{reduce}. 
We now discuss how to reduce the synchronization costs with fine-grained communication.

\section{Fine-grained synchronization}
\label{sec:async-notify-ack}

Barrier based synchronization is an important and widely used operation for synchronizing parallel machine learning programs across multiple machines. After executing a \codesm{barrier} operation, a parallel worker waits until {\em all} the  processes in the system have reached a \codesm{barrier}. Parallel computing libraries like MPI, as well as data parallel frameworks such as BSP systems and some parameter servers expose this primitive to the developers~\cite{chen2015mxnet, gonzalez2012powergraph, malewicz2010pregel, li2014parameterserver}. Furthermore, machine learning systems based on map-reduce use the stage barrier between the \codesm{map} and \codesm{reduce} tasks to synchronize intermediate outputs across machines~\cite{chu2007map, ghoting2011systemml}.


\begin{figure}
\centering
\includegraphics[keepaspectratio=true,width=3.2in]{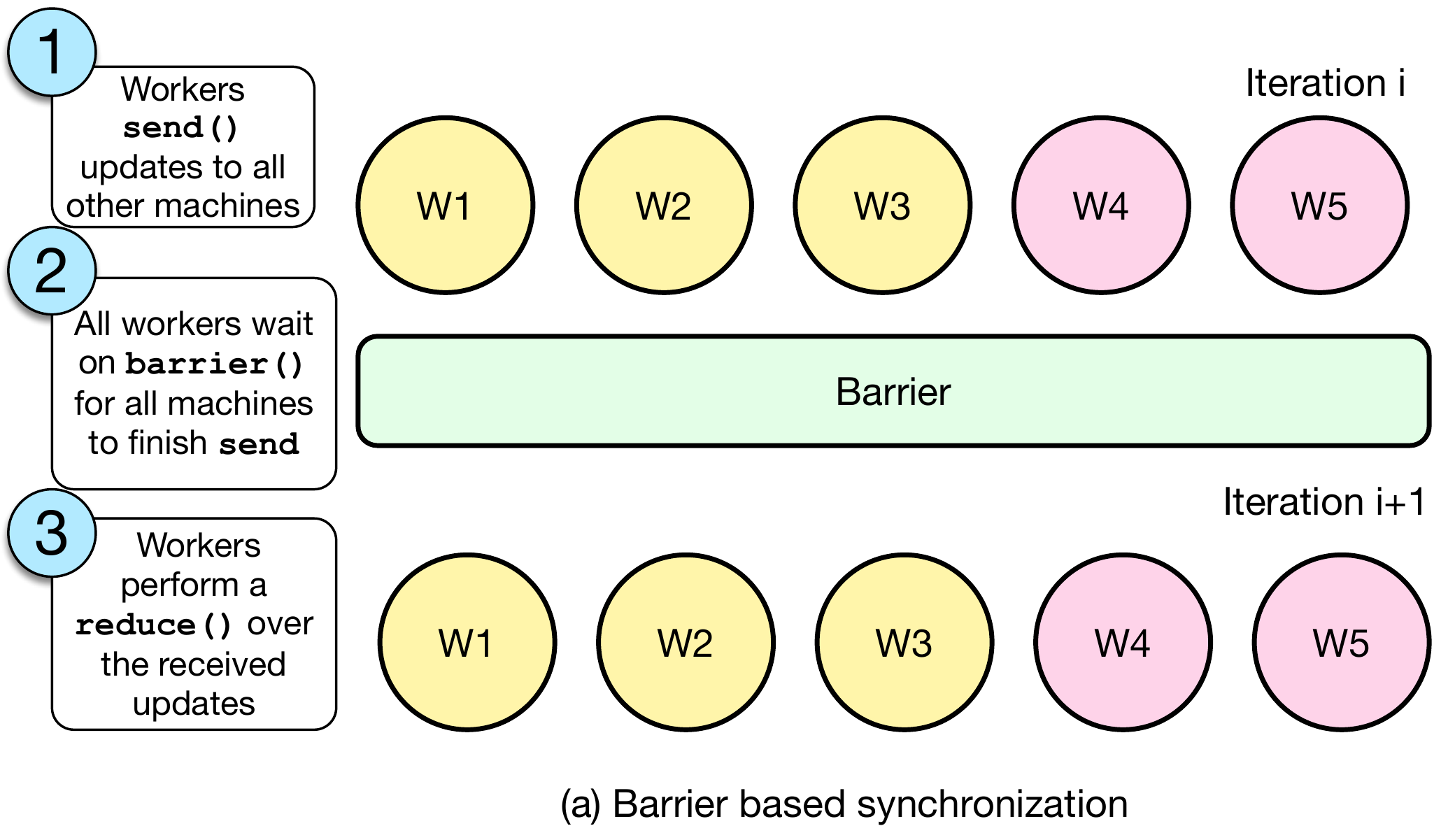}
\caption{\small \bf This figure shows the workers and sparse synchronization semantics. Workers W1, W2 and W3 synchronize with one another. Additionally, workers W3, W4 and W5 synchronize with one another. With a \codesm{barrier}, {\em all} workers wait for every other worker at the barrier and then proceed to the next iteration. This has high performance overhead and may not even guarantee consistency in absence of additional synchronization at each machine.}
\label{fig:barrier}
\vspace{-0.2in}
\end{figure}

Figure~\ref{fig:barrier} shows a parallel training system on $n$ processes. Each process trains on a subset of data, and computes new model parameter values and sends it to all other machines and the waits on the \codesm{barrier} primitive. When all processes arrive at the barrier i.e. all other machines issue a \codesm{send} with their intermediate model parameters, the workers perform a \codesm{reduce} operation over the incoming model parameters and continue processing more input data.

However, using the \codesm{barrier} as a synchronization point in the code suffers from several problems: First, the
BSP protocol described above, suffers from mixed-version issues i.e. in the absence of additional synchronization or serialization at the
receive side, a receiver may perform a \codesm{reduce} with partial or torn model updates (or skip them if a consistency check is enforced). 
This is because just using a \codesm{barrier} gives no information if the recipient has finished receiving and consuming the model update.  Second, most \codesm{barrier} 
implementations synchronize with all other processes in the computation. In contrast, with \reducename, finer grained synchronization primitives are
required that will block on only the required subset of workers to avoid unnecessary synchronization costs. A global \codesm{barrier} operation
is slow and removing this operation can reduce synchronization costs, but makes the workers process the input data at different speeds that may slow
down the overall time to achieve the final accuracy. Finally, using a \codesm{barrier} can cause network resource spikes if all the processes send 
their parameters at the same time.

\begin{figure}
\centering
\includegraphics[keepaspectratio=true,width=3.2in]{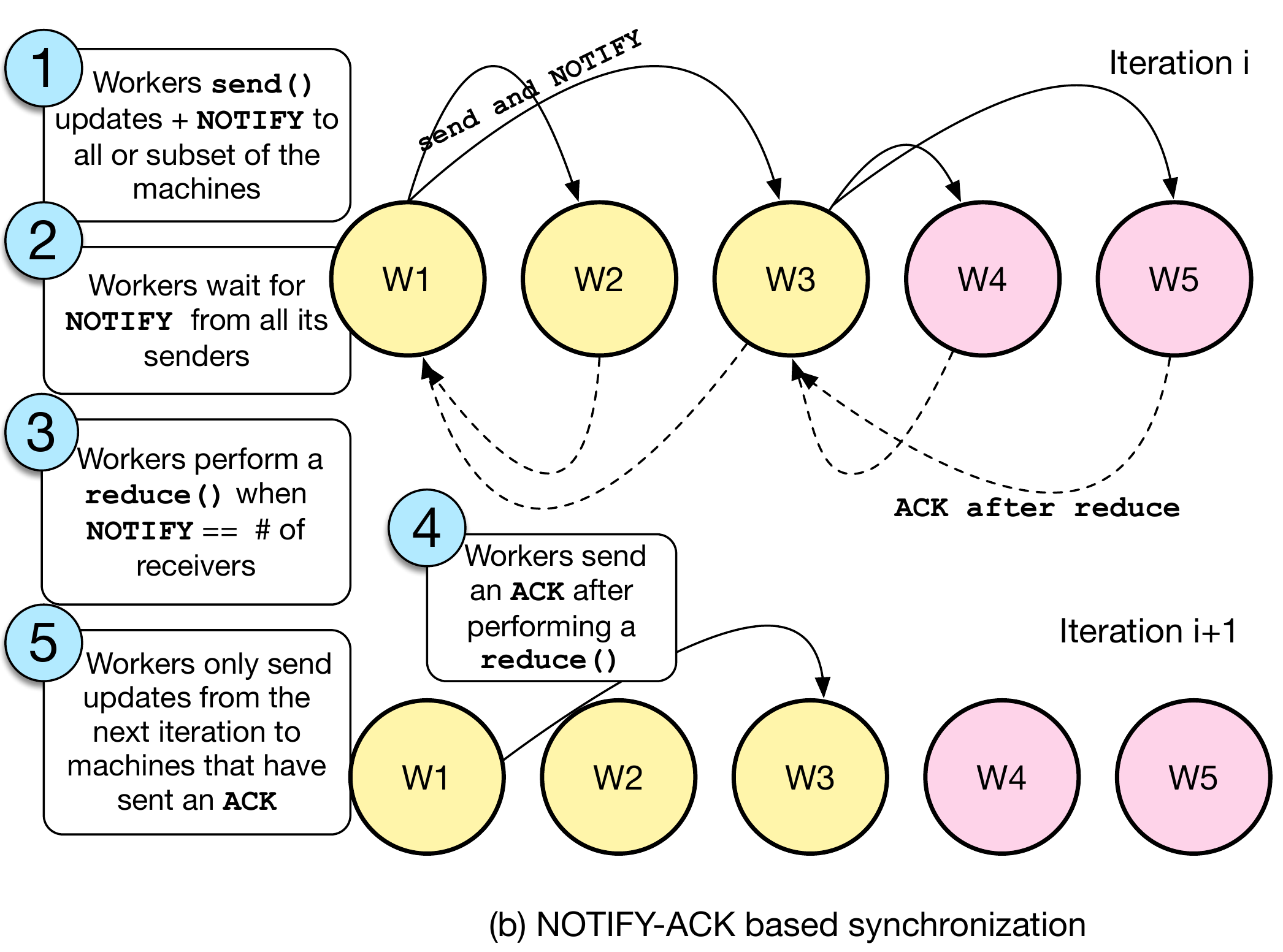}
\caption{\small \bf This figure shows fine-grained synchronization in \toolshort. The solid lines show \codesm{NOTIFY} operation and the dotted lines show the corresponding \codesm{ACK} . Workers only wait for intermediate outputs from dependent workers to perform a \codesm{reduce}. After \codesm{reduce}, workers push more data out when they receive an \codesm{ACK} from receivers signaling that the sent parameter update has been consumed.}
\label{fig:notify-ack}
\vspace{-0.2in}
\end{figure}

Adding extra barriers before/after push and reduce, does not produce a strongly consistent BSP that can incorporate model updates from all replicas since the actual send operation may be asynchronous and there is no guarantee the receivers receive these messages when the perform a reduce. Unless a blocking receive is added after every send, the consistency is not guaranteed. However, this introduces a significant synchronization overhead.

Hence, to provide efficient coordination among parallel model replicas, we require the following three properties from any synchronization
protocol. First, the synchronization should be {\em fine-grained}. Coarse-grained synchronization such as \codesm{barrier} impose high overheads
as discussed above. Second, the synchronization mechanism should provide {\em consistent} intermediate outputs. Strong consistency methods
avoid torn-reads and mixed version parameter vectors, and improve performance~\cite{chen2016revisiting, yun2014nomad}. Finally, the synchronization should be {\em efficient}.
Excessive receive-side synchronization for every \codesm{reduce} and \codesm{send} operation can significantly increase blocking times.

In data-parallel systems with \codesm{barrier} based synchronization, there is often no additional explicit synchronization between the sender and receiver when an update arrives.
Furthermore, any additional synchronization may reduce the performance especially when using low latency communication hardware such as RDMA
that allow one-sided writes without interrupting the receive-side CPU~\cite{kalia2014using}. In the absence of
synchronization, the rate at which the intermediate outputs arrive varies. As a result, a fast sender can overwrite the receive buffers or the receiver may perform a \codesm{reduce} with
a fewer of senders instead of consuming every worker's intermediate output hurting convergence. 

Naiad, a dataflow based data-parallel system, provides a \codesm{notify} mechanism to inform the receivers about the incoming model updates~\cite{murray2013naiad}. This ensures that
when a node performs a local \codesm{reduce}, it consumes the intermediate outputs from all machines. Hence, a per-receiver notification allows for finer-grained synchronization. However, simply using a \codesm{notify} is not enough since a fast sender can overwrite the receive queue of the receiver and a \codesm{barrier} or any other style of additional synchronization is required to ensure that the parallel workers process 
incoming model parameters at the same speeds.


To eliminate the \codesm{barrier} overheads for \reducename\ and to provide strong consistency, we propose using a \codesm{NOTIFY-ACK} based synchronization
mechanism that gives stricter guarantees than using a coarse grained \codesm{barrier}. This can also improve convergence times in some cases since 
it facilitates using consistent data from dependent workers during the \code{reduce} step. 

In \toolshort, with \codesm{NOTIFY-ACK}, the parallel workers compute and send their model parameters with notifications 
to other workers. They then proceed to \codesm{wait} to receive notifications from all its senders as defined by their node 
communication graphs as shown in figure~\ref{fig:notify-ack}. The \codesm{wait} operation counts the \codesm{NOTIFY} events and invokes
the \codesm{reduce} when a worker has received notifications from all its senders as described by the node communication graph.
Once all notifications have been received, it can perform a consistent \codesm{reduce}.

After performing a \codesm{reduce}, the worker sends an \codesm{ACK}, indicating that the intermediate output in previous iteration has been
consumed. Only when a worker receives an \codesm{ACK} for a previous send, indicating that the receiver has consumed the previously sent data,
the worker may proceed to send the data for the next iteration. Unlike a \codesm{barrier} based synchronization, where there is no guarantee 
that a receiver has consumed the intermediate outputs from all senders, waiting on \codesm{ACK}s from receivers ensures that a sender
never floods the receive side queue and avoids any mixed version issues from overlapping intermediate outputs. Furthermore, 
fine-grained synchronization allows efficient implementation of \reducename\ since each sender is only blocked by dependent workers and other workers may run asynchronously. 

\codesm{NOTIFY-ACK} provides clean synchronization semantics in few steps. Furthermore, it requires no additional receive-side synchronization making it ideal for direct-memory access style protocols such as RDMA or GPU Direct~\cite{gpudirect:web}. 
However,  \codesm{NOTIFY-ACK} requires ordering guarantees of the underlying implementation to guarantee that a \codesm{NOTIFY} arrives after the actual data.
Furthermore, in a \codesm{NOTIFY-ACK} based implementation, the framework should ensure that the workers send their intermediate updates and then \codesm{wait} on their reduce inputs 
to avoid any deadlock from a cyclic node communication graphs.

\section{Implementation}
\label{sec:impl}

We develop our second generation distributed learning framework using the \toolshort\ model to incorporate \reducename\ and fine-grained synchronization. We implement distributed data-parallel model averaging over stochastic gradient descent (SGD). We implement our reference framework with \reducename\ and \codesm{NOTIFY-ACK} support in C++ and provide Lua bindings to run existing Torch and RAPID deep learning networks~\cite{collobert2011torch7, milde:web}. 

For distributed communication, we use MPI and create the model parameters in distributed shared memory. In our implementation, parallel model replicas create a model vector in the shared memory and train on a portion of the dataset using the SGD algorithm. To reduce synchronization overheads, each machine maintains a per-sender receive queue to receive the model updates from other machines~\cite{power2010piccolo}. The queues and the shared memory communication between the parallel model replicas are created based on a node communication graph provided as an input when launching a job. Periodically, after processing few data examples and updating the local model $w$, the parallel replicas communicate the model parameters. The model replicas then average the incoming parameters with their own local model vector($w$). 

We use the infiniBand transport and each worker directly writes the intermediate model to its senders without interrupting the receive side CPU, using one-sided RDMA operations. After the \codesm{reduce} operation, each machine sends out the model updates to other machines' queues as defined by the communication graph. Our system can perform \codesm{reduce} over any user-provided node communication graph allowing us to evaluate \reducename\ for different node communication graphs. 

Furthermore, we also implement the synchronous, asynchronous and \codesm{NOTIFY-ACK} based synchronization. We implement synchronous (BSP) training by using the MPI provided \codesm{barrier} primitives. We use low-level distributed \codesm{wait} and \codesm{notify} primitives to implement \codesm{NOTIFY-ACK}. We maintain ACK counts at each node and send all outputs before waiting for ACKs across iterations to avoid deadlocks.

 We use separate infiniBand queues for transmitting short messages (ACKs and other control messages) and large model updates (usually a fixed size for a specific dataset). For Ethernet based implementation, separate TCP flows can be used to reduce the latency of control messages~\cite{nightingale2012flat}. We provide fault tolerance by check-pointing the trained model periodically to the disk. Additionally, for synchronous methods, we implement a dataset specific dynamic timeout which is a function of the time taken for the \codesm{reduce} operation.

\section{Evaluation}
\label{sec:evaluation}
In this section, we evaluate the \toolshort\ model using the following criterion.

\begin{itemize}

\item What is the benefit of \reducename? Do networks with a higher spectral gap exhibit better convergence?

\item What is the benefit of using fine-grained synchronization over a \codesm{barrier}? Is it consistent?

\end{itemize}

\begin{figure}
\centering
\includegraphics[keepaspectratio=true,width=3.2in]{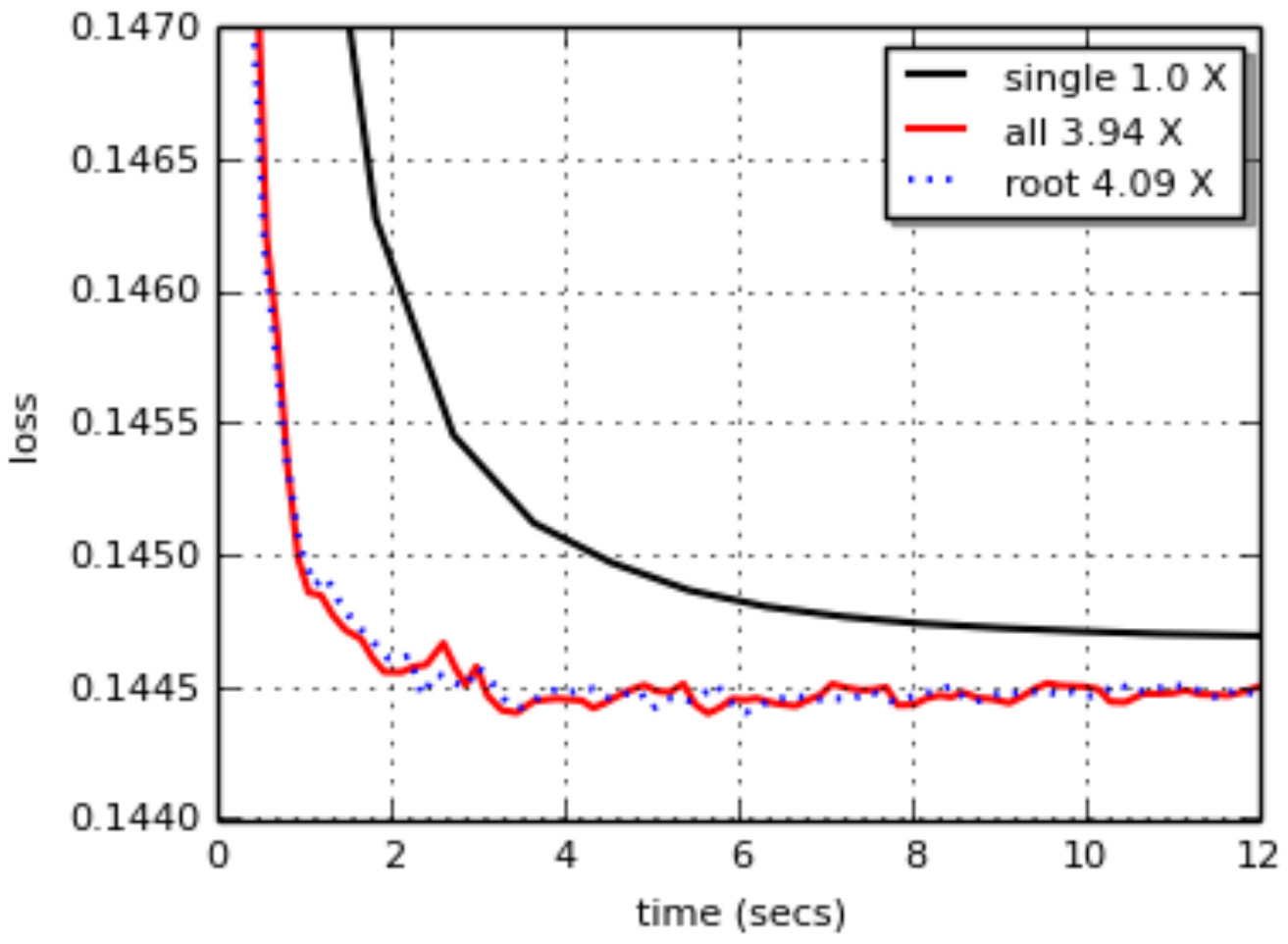}
\caption{\small \bf This figure compares the convergence of a root expander graph with an all-reduce graph over a single machine SGD. Each machine in root expander graph transmits 56 MB data while all-reduce transmits 84 MB of data to reach the desired accuracy value. Speedups are measured over a single-machine convergence.}
\label{fig:ap-speedup}
\vspace{-0.2in}
\end{figure}

We evaluate the \toolshort\ model for applications that are commonly used today including text classification, spam classification, image classification and Genome detection. Our baseline for evaluation is our BSP/synchronous and asynchronous based all-reduce implementations which is provided by many other distributed big data or machine learning systems~\cite{abaditensorflow, apachehama:web, dai2013petuum, gabriel2004open, gonzalez2012powergraph, gregor2005parallel, malewicz2010pregel,power2010piccolo}. We use an efficient infiniBand implementation stack that improves performance for all methods. However, \reducename\ and \codesm{NOTIFY-ACK} can be implemented and evaluated over any existing distributed learning platform such as GraphLab or TensorFlow. 

We run all our experiments on eight Intel Xeon 8-core, 2.2 GHz Ivy-Bridge processors and 64 GB DDR3 DRAM. All  connected via a Mellanox Connect-V3 56 Gbps infiniBand cards to an infiniBand backplane. Our 56 Gbps infiniBand network architecture provides a peak throughput of slightly over 40 Gbps after accounting for the bit-encoding overhead for reliable transmission. All machines load the input data from a shared NFS partition. We run multiple processes, across these machines and run multiple processes on each machine, especially for models with less than 1M parameters, where a single model replica is unable to saturate the network and CPU. All reported times do not account for the initial one-time cost for the loading the data-sets in memory. All times are reported in seconds. 

We evaluate the two machine learning methods:

\begin{enumerate}
\item SVM: We test \toolshort\ on distributed SVM based on Bottou's SVM-SGD~\cite{bottou-2010}. Each machine computes the model parameters and communicates them to other machines as described in the machine communication graph. We train SVM over the RCV1 dataset (document classification), the webspam dataset (webspam detection) and the splice-site dataset (Genome classification)~\cite{pascal:web}.

\item Convolutional Neural Networks (CNNs): We train CNNs for image classification over the CIFAR-10 dataset~\cite{cifar10:web}. The dataset consists of 50K train and 10K test images and the goal is to classify an input image to one amongst 10 classes. We use the VGG network to train 32x32 CIFAR-10 images with 11 layers that has 7.5M parameters~\cite{simonyan2014very}. We run data parallel replicas on multiple machines. We use \codesm{OMP\_PARALLEL\_THREADS} to parallelize the convolutional operations within a single machine.

\end{enumerate}

\subsection{Approximate processing benefits}

\begin{figure}
\centering
\includegraphics[keepaspectratio=true,width=3.2in]{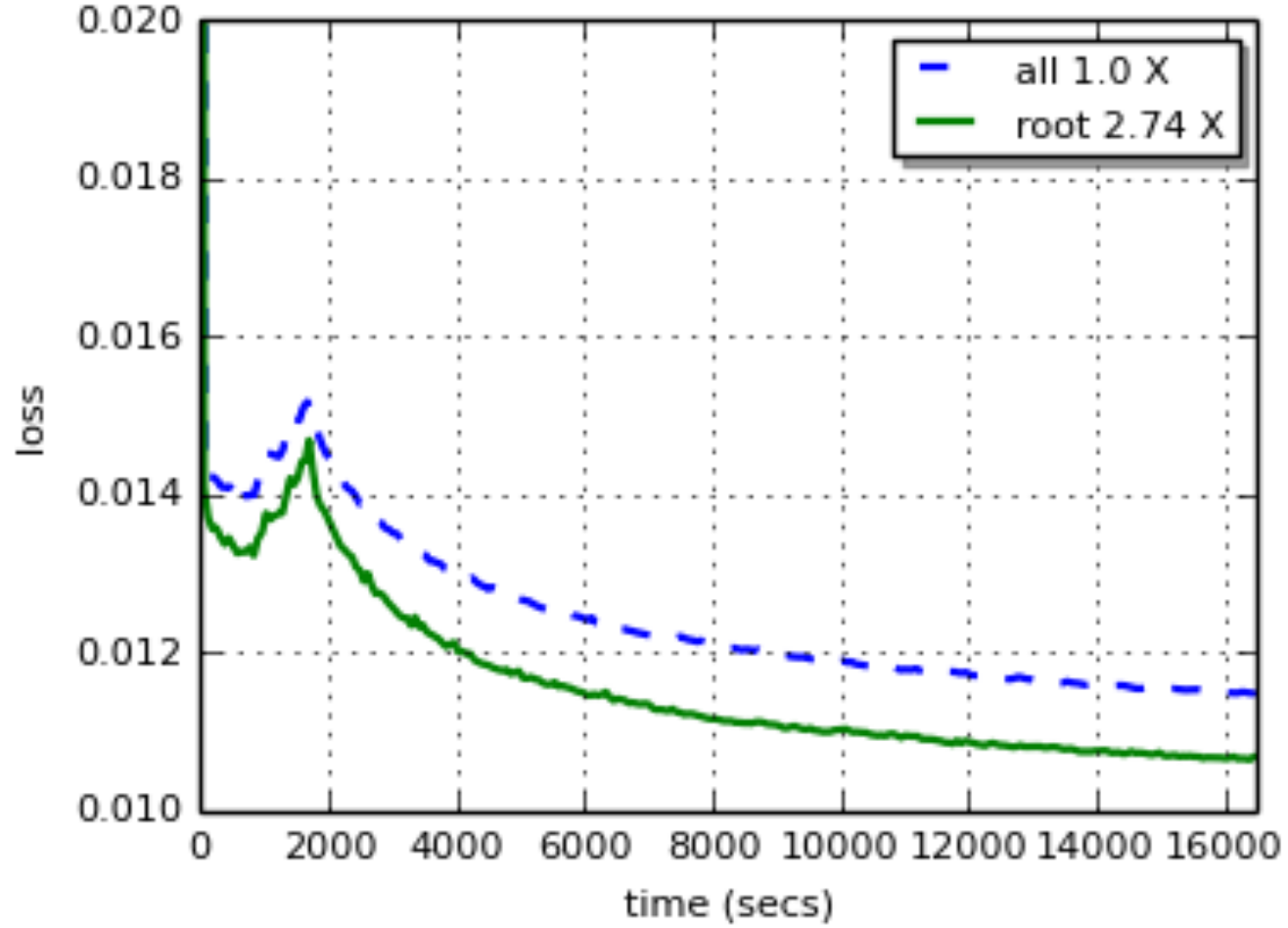}
\caption{\small \bf This figure shows the convergence of a root expander graph with all-reduce graph for the splice-site dataset. Each machine in root graph transmits 219GB data/process while all-reduce transmits 2.08TB/worker of data to reach the desired accuracy. Speedups are measured over all-reduce.}
\label{fig:ap-speedup2}
\vspace{-0.2in}
\end{figure}


\paragraph {Speedup with \reducename} We measure the speedups of all applications under test as the time to reach a specific accuracy. We first evaluate a small dataset (RCV1, 700MB, document classification) for the SVM application. The goal here is to demonstrate that for problems that fit in one machine, our data-parallel system outperforms a single thread for each workload~\cite{mcsherry2015scalability}. Figure ~\ref{fig:ap-speedup} shows the convergence speedup for 4 machines for the RCV1 dataset. We compare the performance for all-reduce against a root graph with a fixed out degree of two, where each node sends the model updates to  two nodes -- its neighbor and $rootN th$ node. We find that for the RCV1 dataset, the root expander graph converges marginally faster than the all-reduce primitive, owing to lower network costs of sending the intermediate model and lower CPU costs of processing fewer incoming models at each of the machines.


Figure~\ref{fig:ap-speedup2} shows the convergence for the SVM application using the splice-site dataset on 8 machines with $8$ processes. The splice site training dataset is about 250GB, and does not fit in-memory in any one of our machines. This is one of the largest public dataset that cannot be sub-sampled and requires training over the entire dataset to converge correctly~\cite{agarwal2014reliable}. Figure~\ref{fig:ap-speedup2} compares the two  communication graphs -- an all-reduce graph and a root expander graph with an out-degree of $2$. We see that the expander graph can converge faster, and requires about 10X lower network bandwidth. Hence, \reducename\ can improve convergence time and reduce network bandwidth requirement as compared to a na\"{\i}ve all-reduce.

\begin{figure}
\centering
\includegraphics[keepaspectratio=true,width=3.2in]{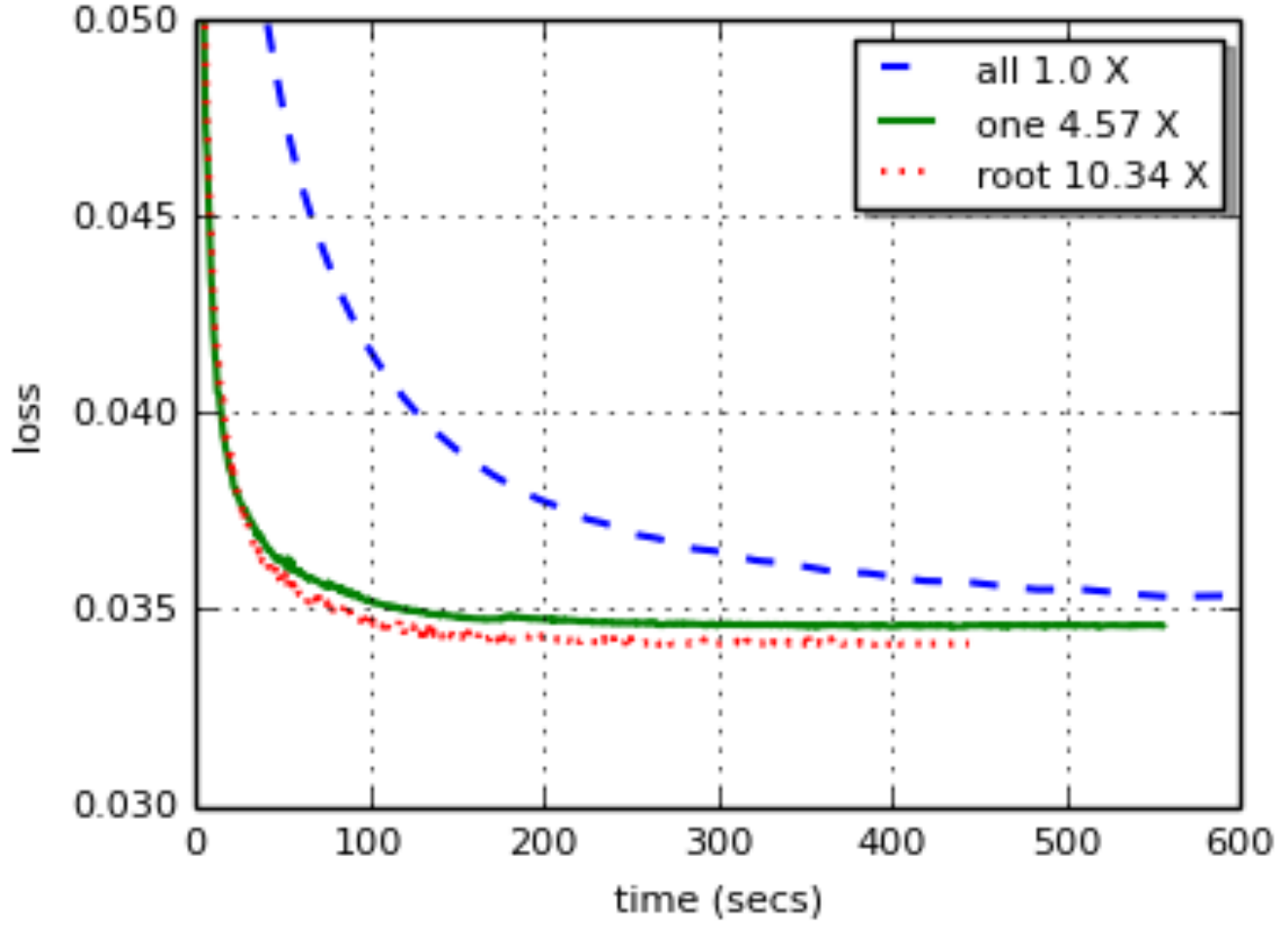}
\caption{\small \bf This figure shows the convergence of a chain (one) graph and a root expander graph as compared to an all-reduce implementation for the webspam dataset over 25 workers.}
\label{fig:ap-speedup-graphs}
\vspace{-0.2in}
\end{figure}


Figure ~\ref{fig:ap-speedup-graphs} shows the convergence for the SVM application on $25$ processes using the webspam dataset consisting of 250K examples. The data is split across all processes and each machine only trains over 31.25K training set examples. We compare three node communication graphs -- a root expander graph (with a spectral gap of $0.2$ for $25$ nodes) and a (one) chain-like architecture where each machine maintains a model and communicates its updates in the form a chain to one other machine, with the all-reduce implementation. The chain node graph architecture has lower network costs, but very low spectral values (around 0.008).  Hence, it converges slower than the root expander graph. Both the sparse graphs provide a speedup over all-reduce since the webspam dataset has a large dense parameter vector of about 11M \codesm{float} values. However, the chain graph requires more epochs to train over the dataset and sends 50433 MB/node to converge while the root node requires 44351 MB/node even though the chain graph transmits less data per epoch. Hence, one should avoid sparse reduce graphs with very low spectral gap values and use expander graphs provide good convergence with reasonable network costs.

To summarize, we find that \reducename\ can provide significant speedup in convergence (by 2-10X) and reduces the network and CPU costs. However, if the node communication graph is sparse and has low spectral gap values (usually less than $0.01$), the convergence can be slow. Hence, \reducename\ provides a quantifiable measure using the spectral gap values to sparsify the node communication graphs. For models where the network capacity and CPU costs can match the model costs, using a network that can support the largest spectral gap is recommended.

\subsection{Fine-grained synchronization benefits}

We evaluate the benefits of fine-grained synchronization in this section. We compare the performance of \codesm{NOTIFY-ACK}, synchronous and asynchronous implementation. We implement the BSP algorithm using a \codesm{barrier} in the training loop, and all parallel models perform each iteration concurrently. However, simply using a \codesm{barrier} may not ensure consistency at the receive queue. For example, the parallel models may invoke a \codesm{barrier} after sending the models and then perform a \codesm{reduce}. This does not guarantee that each machine receives {\em all} the intermediate outputs during reduce. Hence, we perform a consistency check on the received intermediate outputs and adjust the number of intermediate models to compute the model average correctly. Each parameter update carries a unique version number in the header and footer we verify the version values in the header and footer are identical before and after reading the data from the input buffers.

For asynchronous processing, we perform no synchronization between the workers once they finish loading data and start training. We check the incoming intermediate model updates for consistency as described above. For the \codesm{NOTIFY-ACK} implementation, we send intermediate models with notifications and \codesm{wait} for the ACK before performing a \codesm{reduce}. Some network interconnects (or libraries over these interconnects) may not guarantee that the notifications arrive with the data, and we additionally check the incoming model updates and adjust the number of reducers to compute the model average correctly. We first perform a micro-benchmark to measure how many consistent \codesm{reduce} operations do synchronous (SYNC), asynchronous (ASYNC) and \codesm{NOTIFY-ACK} style of synchronization provide. 

\begin{figure}
\centering
\includegraphics[keepaspectratio=true,width=3.2in]{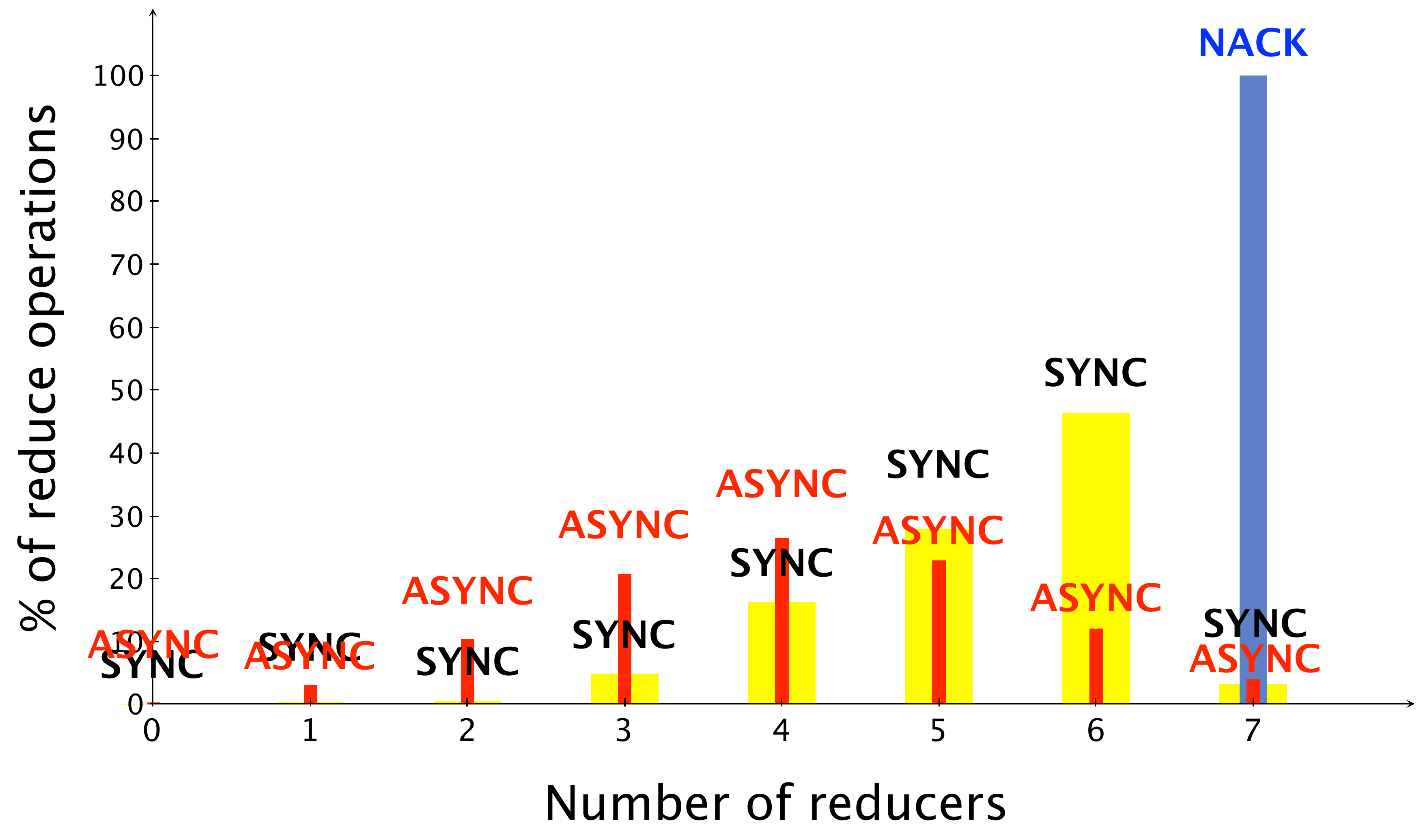}
\caption{\small \bf This bar graph shows the percentage of consistent \codesm{reduce} operations with \codesm{NOTIFY-ACK} vs BSP vs ASYNC for $8$ machines for the RCV1 dataset. \codesm{NOTIFY-ACK} provides the strongest consistency allowing 100\% of \codesm{reduce} operations to be performed using other $7$ nodes.}
\label{fig:as-reducecounts}
\vspace{-0.2in}
\end{figure}

Figure~\ref{fig:as-reducecounts} shows, for a graph of 8 nodes with each machine having an in-degree 7 , the distribution of correct buffers reduced. \codesm{NOTIFY-ACK}, reduces with all 7 inputs and is valid 100\% of the time. BSP has substantial torn reads, and 77\% of the time performs a \codesm{reduce} with 5 or more workers. ASYNC can only perform 39\% of the \codesm{reduce} operations correctly with 5 or more workers. Hence, we find that \codesm{NOTIFY-ACK} provides the most consistent data during \codesm{reduce} and fewest torn buffers followed by BSP using a \codesm{barrier}, followed by asynchronous. We now evaluate the benefits of \codesm{NOTIFY-ACK} with all-reduce communication.

\begin{figure}
\centering
\includegraphics[keepaspectratio=true,width=3.2in]{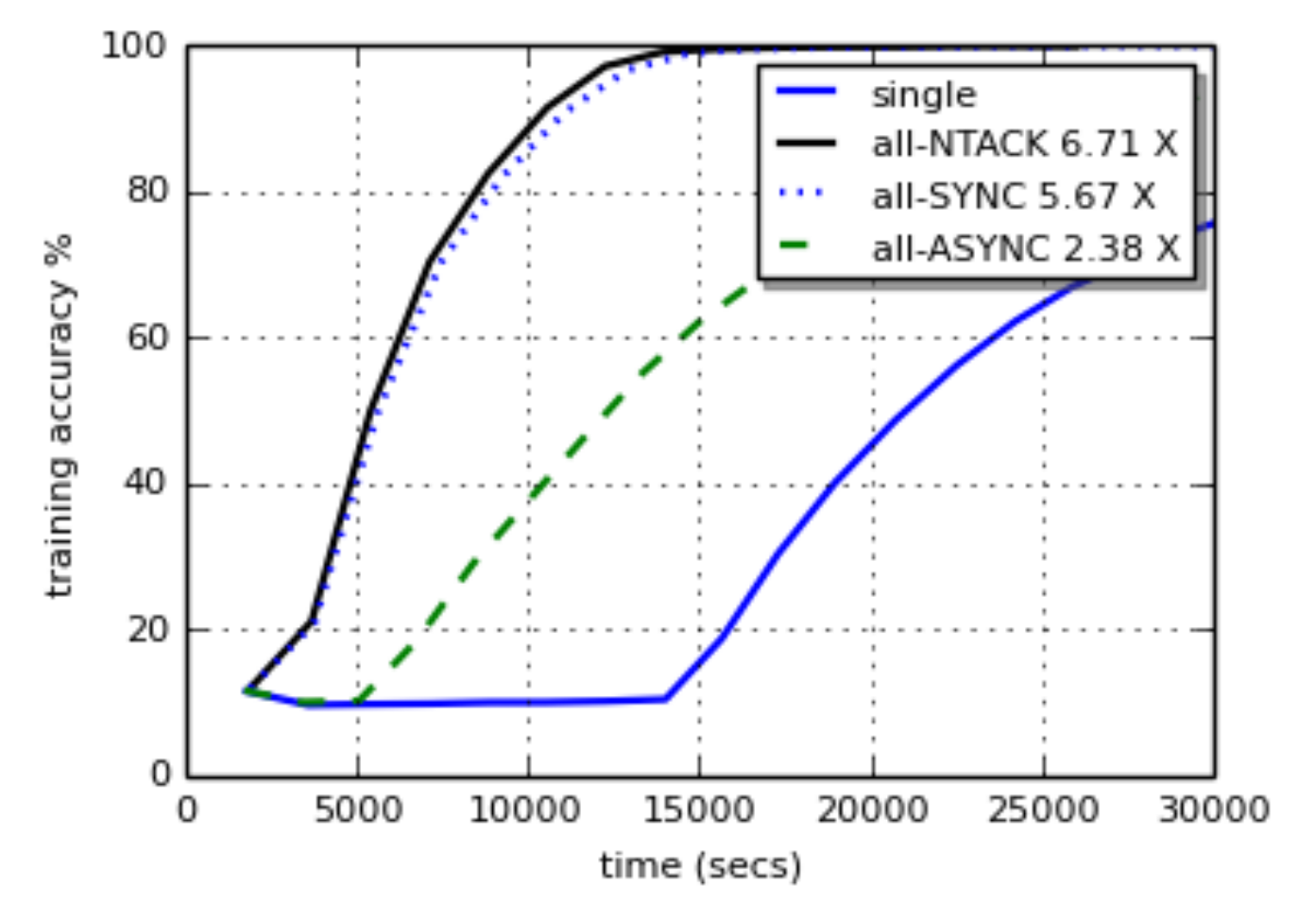}
\caption{\small \bf This figure shows the convergence of \codesm{NOTIFY-ACK}, BSP, ASYNC with eight machines using all-reduce for training CNNs over the CIFAR-10 dataset with a VGG network. Speedups are measured over a single machine implementation.}
\label{fig:as-speedup2}
\vspace{-0.2in}
\end{figure}

Figure~\ref{fig:as-speedup2} shows the convergence for the CIFAR-10 dataset for eight machines in an all-reduce communication. We calculate the time to reach 99\% training accuracy on the VGG network which corresponds to an approximately 84\% test accuracy. We train our network with a mini-batch size of 1, with no data augmentation and a network wide learning rate schedule that decays every epoch. We find that with CNNs, \codesm{NOTIFY-ACK} provides superior convergence over BSP and ASYNC. Even with a dense communication graph of all-reduce, we find that \codesm{NOTIFY-ACK} reduces \codesm{barrier} times and provides stronger consistency than BSP providing competitive throughput and fast convergence. Furthermore, we find that ASYNC initializes slowly and converges slower than synchronous and \codesm{NOTIFY-ACK} methods. 

We also measure the throughput (examples/second processed) for the three different synchronization methods -- synchronous (BSP), asynchronous and the \codesm{NOTIFY-ACK} method. With \codesm{NOTIFY-ACK}, we avoid coarse-grained synchronization and achieve a throughput of 229.3 frames per second (or images per second) for eight machines. This is the average processing time and includes the time for forward and backward propagation, adjusting the weights and communicating and averaging the intermediate updates. With BSP, we achieve 217.8 fps. Finally, we find that even though ASYNC offers the highest throughput of on an average 246 fps, Figure ~\ref{fig:as-speedup2} shows that the actual convergence is poor. Hence, to understand the benefits of approaches like relaxed consistency, one should consider speedup towards a (good) final accuracy apart from throughput. However, there are some cases where ASYNC may provide good performance when the communication between the workers is not crucial for convergence. This may happen when the dataset is highly redundant. Second, if the model is sufficiently sparse which makes the \codesm{reduce} operation commutative and reduces conflicting updates.

\begin{figure}
\centering
\includegraphics[keepaspectratio=true,width=3.2in]{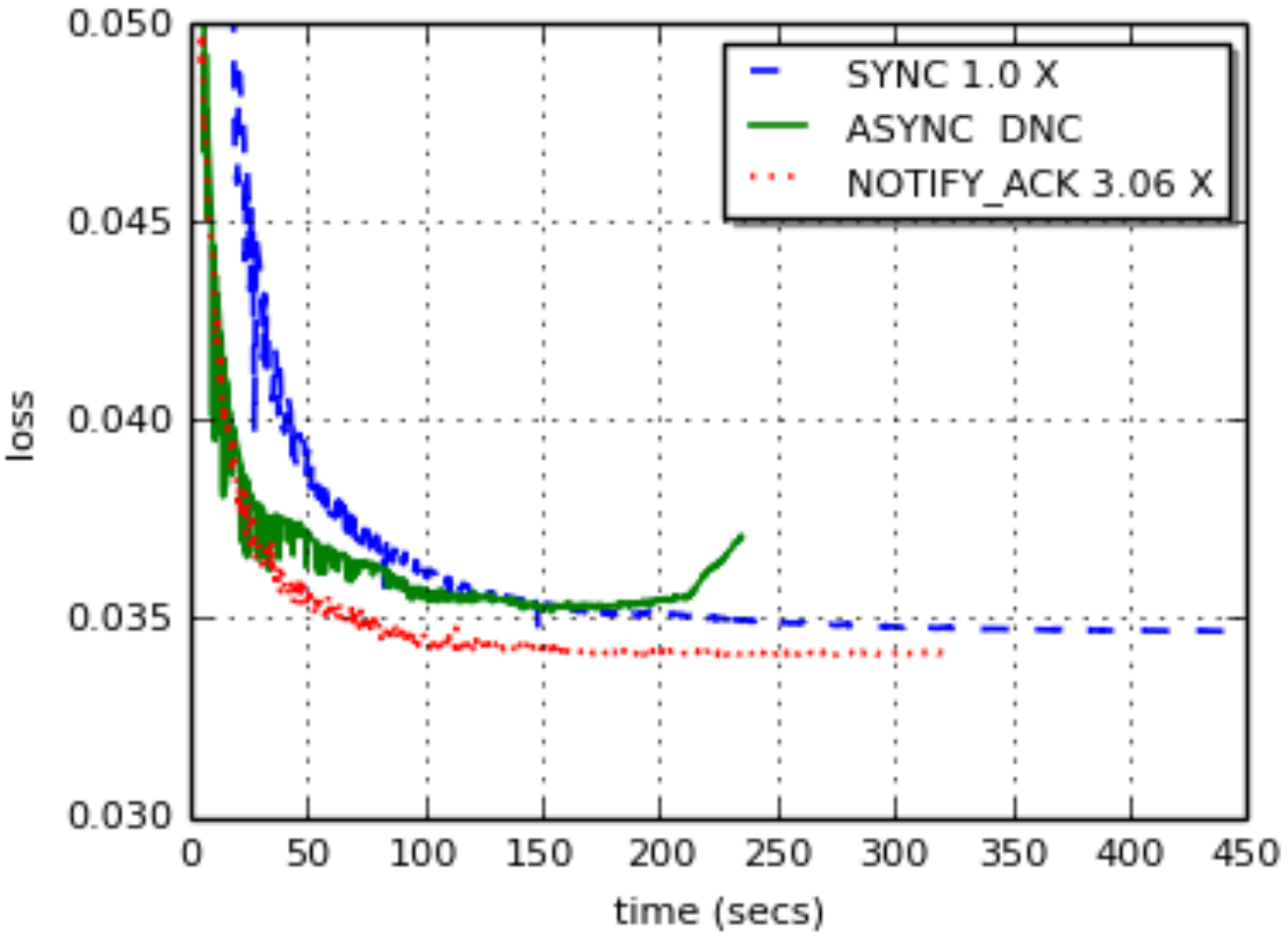}
\caption{\small \bf This figure shows the convergence of \codesm{NOTIFY-ACK} vs BSP and ASYNC for the root expander graph over the webspam dataset. The asynchronous implementation does not converge (DNC) to the final value.}
\label{fig:as-speedup}
\vspace{-0.05in}
\end{figure}

Figure ~\ref{fig:as-speedup} shows the convergence for the SVM application using the webspam dataset on $25$ processes. We use the root-expander graph as described earlier, where each node receives the model updates with two other nodes. We find that using fine-grained \codesm{NOTIFY-ACK} improves the convergence performance and is about $3X$ faster than BSP for the webspam dataset. Furthermore, the asynchronous algorithm does not converge to the correct value even though it operates at a much higher throughput. \codesm{NOTIFY-ACK} provides good performance for three reasons. First, \codesm{NOTIFY-ACK} provides stronger consistency than BSP implemented using a barrier. In the absence of additional heavy handed synchronization with each sender, the model replicas may \codesm{reduce} with fewer incoming model updates. \codesm{NOTIFY-ACK} provides stronger guarantees since each worker waits for the \codesm{NOTIFY}s before performing the reduce and sends out additional data with after receiving the \codesm{ACK} messages. Second, for approximate processing i.e. when the communication graph is sparse, a \codesm{barrier} blocks all parallel workers while when using fine-grained communication with \codesm{NOTIFY-ACK} independent workers run asynchronously with respect to one another. 

To summarize, we find that \codesm{NOTIFY-ACK} eliminates torn buffers and provides stronger consistency over BSP and asynchronous communication for dense as well as sparse node graphs. We describe our results for the BSP/all-reduce model. However, these results can also be extended to bi-directional communication architectures such as the parameter server or the butterfly architecture.

\section{Related Work}
\label{sec:related}
We discuss related work about batch processing systems, data-parallel approximate and
asynchronous processing and distributed machine learning frameworks.

The original map-reduce uses a stage barrier i.e. all mapper tasks synchronize intermediate outputs over a distributed file-system~\cite{dean2008mapreduce}. 
This provides synchronization and fault tolerance for the intermediate job states but can affect the job performance due to frequent disk I/O. 
Spark~\cite{zaharia2012resilient} implements the map-reduce model in-memory using copy-on-write data-structures (RDDs). 
The fault-tolerance is provided by checkpoint and replay of these RDDs which restrict applications from running asynchronously.
ASIP~\cite{gonzalez2015asynchronous} adds asynchrony support in Spark by decoupling fault tolerance and coarse-grained synchronization and incorporates fully asynchronous execution in Spark. However, ASIP finds that asynchronous execution may lead to incorrect convergence, and presents the case for running asynchronous machine learning jobs using second order methods that provides stronger guarantees but can be extremely CPU intensive~\cite{ouyang2013stochastic}. Finally, there are many existing general purpose dataflow based systems that use barriers or block on all inputs to arrive~\cite{condie2010mapreduce, isard2007dryad, murray2013naiad, olston2008pig}. 
\toolshort\ uses fine-grained synchronization with partial reduce to mitigate communication and synchronization overheads. 
Fault tolerance can be incorporated in the \toolshort\ model by using application checkpoints as described in our example implementation.

Past work on partial aggregation has proposed efficient tree-style all-reduce primitive to mitigate communication costs. 
This is extensively used in batch systems to reduce network costs by combining results 
at various levels such as machine, rack etc.~\cite{chaiken2008scope, yu2008dryadlinq}.
However, the \codesm{reduce} operation suffers from additional latency proportional 
to the height of the tree. Furthermore, when partial aggregation is used with iterative-convergent 
algorithms, the workers wait for a significant aggregated latency time which can be undesirable. Other work on partial-aggregation 
produces variable accuracy intermediate outputs over different computational budgets~\cite{kumar2016hold, shatdal1995adaptive}.
\toolshort\ exploits the iterative nature of machine learning algorithms for sparse aggregation of intermediate outputs 
that are propagated indirectly to all other workers over {\em successive} iterations reducing synchronization costs. \toolshort\
provides the same final accuracy as an aggregated all-reduce over successive iterations. The workers 
communicate over a network topology designed to disperse the intermediate outputs over fewer iterations. Other methods to reduce network costs include using lossy compression~\cite{abaditensorflow} or KKT filters~\cite{li2014parameterserver}. The latter method determines the utility of intermediate updates are before sending them over the network. These methods can be applied with \reducename\ even though they may incur additional CPU costs unlike \reducename. 

Past work has explored removing barriers in Hadoop to start reduce operations as soon as some of the mappers finish execution~\cite{goiri2015approxhadoop, verma2013breaking}. HogWild~\cite{recht2011hogwild} provides a single shared parameter vector and allows parallel threads to update model parameters without locks thrashing one another's updates. However, HogWild may
not converge to a correct final value if the parameter vector is dense
and the updates from different machines overwrite one-another frequently. Project Adam~\cite{chilimbi2014adam}, DogWild~\cite{noel2014dogwild} and other systems that use HogWild
in a distributed setting results in excessive wasted communication especially when used to communicate dense parameter updates. A large number of previous systems propose removing the barriers which may provide faster throughput but may lead to slower or incorrect convergence towards the final optimization goal. To overcome this problem, bounded-staleness~\cite{cui2014exploiting} provides asynchrony within bounds for the parameter server i.e. the forward running threads wait for the stragglers to catch up. However, determining these bounds empirically is difficult, and in some cases they may not more relaxed than synchronous. \toolshort\ instead proposes using fine-grained synchrony that reduces synchronization overhead and provides stronger guarantees than what can be provided through barriers. 

There has been a flurry of distributed machine learning platforms developed recently. Parameter Server~\cite{dean2012large, li2014parameterserver} provide a master-slave style communication framework. Here, workers compute the parameter updates and send it to a central server (or a group of servers). The parameter server computes and updates the global model and sends it to the workers and they continue to train new data over this updated model. Hence, in the parameter server the workers need to wait after every batch to receive an updated model. Here, the overall communication is low even though the master has a high-fanout of nodes it communicates with. Furthermore, since the master maintains a global model and resets the stragglers with its own updated model, the parameter server architecture always converges to the correct value. On the contrary, all-reduce based systems, usually implemented over map-reduce or BSP based systems, use a network of workers that synchronize using a \codesm{barrier}. All-reduce may operate fully asynchronously since unlike parameter server there is no consensus operation to exchange the gradients. However, they suffer from high communication costs as the number of workers increase. \toolshort\ reduces communication overheads in the all-reduce model and by proposing partial-reduce based on information dispersal properties of underlying nodes. 

TensorFlow runs a dataflow graph across a cluster. Unlike some other parameter servers that use \codesm{barrier} for synchronous training, TensorFlow uses a queue abstraction and uses the asynchronous parameter server to train large models~\cite{abaditensorflow}. Recently, instead of using asynchronous synchronization methods, Tensorflow has introduced synchronization with backup workers that skips sychnronizing with the tail of the workers~\cite{chen2016revisiting}. Using \toolshort's \codesm{NOTIFY-ACK} style synchronization can provide stronger consistency without any additional workers. Additionally, for large models, the large fanout of the master can be a bottleneck and the model parameters are aggregated at bandwidth hierarchies~\cite{chen2016revisiting}. Using \toolshort's \reducename\ to improve the convergence behavior of such network architectures can reduce the wait times. The parameter server architecture has also been proposed over GPUs~\cite{cui2016geeps,zhang2015poseidon} and the communication and synchronization costs can be reduced in these systems by using the \toolshort\ model. 



\section{Conclusion} \label{sec:conclusion} 

Existing big-data processing frameworks use approximation and asynchrony to
accelerate job processing throughput (i.e. examples/second processed). However,
these optimizations may not benefit the overall accuracy of the job output and
may even result in a slowdown as compared to the bulk- synchronous model.

In this paper, we introduce an asynchronous and approximate processing model
that reduces synchronization costs and provides strong quantifiable guarantees
allowing application developers to reason about the guarantees they provide. In our
results, we demonstrate that a framework written using the \toolshort\ model
provides 2-10X speedups in convergence and up to 10X savings in network costs. Other
optimization problems such as graph-processing face similar trade-offs, and can
benefit from using the \toolshort\ model.

\section*{Acknowledgments}
We would like to thank Cun Mu for his help with the analysis of stochastic reduce convergence, and Igor Durdanovic for helping us port RAPID to MALT-2. Finally, we would like to thank Hans-Peter Graf for his support and encouragement.

\pagebreak\

\begin{appendices}
\section{Stochastic reduce convergence analysis}
\label{sec:proof}

In this section, we provide the supplementary material to analyse the convergence for any optimization algorithm that is implemented using \reducename. Mathematically, the optimization problem is defined on a connected undirected network and solved by $n$ nodes collectively,

\begin{flalign}
\min_{\bm x \in \mc X \subseteq \reals^d}  \quad \bar f(\bm x):= \sum_{i=1}^n f_i(\bm x).
\end{flalign}

The feasible set $\mc X$ is a closed and convex set in $\reals^d$ and is known by all nodes, whereas $f_i: \mc X \in \reals$ is a convex function privately known by the node $i$. We also assume that $f_i$ is $L$-Lipschitz continuous over $\mc X$ with respect to the Euclidean norm $\norm{\cdot}{}$. The network $\mc G = (\mc N, \mc E)$, with the node set $\mc N = [n] := \set{1,2,\cdots, n}$ and the edge set $\mc E \subseteq \mc N \times \mc N$, specifies the topological structure on how the information is spread amongst the nodes through local node interactions over time. Each node $i$ can only send and retrieve information as defined by the node communication graph $\mc N(i):=\set{j \;\vert\; (j,i) \in \mc E}$ and itself. 

In this algorithm, each node $i$ keeps a local estimate $\bm x_i$ and a model variable $\bm w_i$ to maintain an accumulated sub-gradient. At iteration $t$, to update $\bm w_i$, each node needs to collect the model update values of its neighbors which is the gradient value denoted by $\nabla f_i(\bm w_t^i)$, and forms a convex combination with an equal weight of the received information. 
The learning rate is denoted by $\eta_t$. Hence, the updates received by each machine can be expressed as:



\begin{flalign}
\bm w_{t+1/2}^i &\gets \bm w_t^i - \eta_t \nabla f_i(\bm w_t^i) \label{eqn:alg}\\
\bm w_{t+1}^i &\gets \frac{1}{|\mc N_{in}^i|} \sum_{j \in \mc N_{in}^i} \bm w_{t+1/2}^j \nonumber
\end{flalign}

\paragraph{Network requirement.} In order to make the above algorithm work, we need a network over which each node has the same influence. To understand and quantify this requirement, we denote the adjacency matrix as $\mb A$, i.e. $A_{ij} = 1$ if $(j,i) \in \mc A$ and $0$ otherwise, and denote $\mb P$ as the matrix after scaling each $i$-th row of $\mb A$ by the in-degree of node $i$, i.e. $\mb P = \diag{\bm d_{in}}^{-1} \mb A$, where $\bm d_{in} \in \reals^k$ and $d_{in}(i)$ equals the in-degree of node $i$. For ease of illustration, we assume that $d=1$ and $w^i_0 = 0$. Denote $\bm g_t = (g^1_t, g^2_t, \ldots, g_t^k)$, and $\bm w_t = (w^1_t, w_t^2, \ldots, w_t^k)^\top$. Then the updates can be tersely expressed as :
\begin{flalign}
\bm w_1 &= - \eta_0 \mb P \bm g_0 \nonumber \\
\bm w_2 &= - \eta_0 \mb P^2 \bm g_0 - \eta_1 \mb P \bm g_1 \nonumber \\
& \vdots \nonumber\\
\bm w_{t} &= - \eta_0  \mb P^t \bm g_0 - \eta_1 \mb P^{t-1} \bm g_1 -  \cdots - \eta_{t-1} \mb P \bm g_{t-1}  \nonumber\\ 
\bm  w_{t} &= -\sum_{k=0}^{t-1} \eta_k \mb P^{t-k} \bm g_k \label{eqn:ill} 
\end{flalign}
It can be easily verified that $\mb P^\infty:=\lim_{t\to \infty} \mb P^t = \bm 1 \bm \pi^\top,$ where $\bm \pi$ is a probability distribution (known as stationary distribution). Thus, $\pi_i$ here represents the influence that node $i$ played in the network. Therefore, to take a fair treatment of each node, $\pi_i = \frac{1}{k}$ is desired, which is equivalent to saying the row sums and column sums of $\mb P$ are all equal to one, i.e. $\mb P$ is a doubly stochastic matrix. In the context of our network setting, we need a network whose nodes all have the same in-degree. Indeed, when $f_i$ is convex, the convergence results have been established under this assumption \cite{nedic2009distributed}.

\paragraph{Spectral Gap:} 
Besides being regular, the network $\mc N(i)$ should also be constructed in a way such that the information can be effectively spread out over the entire network, which is closely related with the concept of spectral gap. We denote $\sigma_1(\mb P) \ge \sigma_2(\mb P) \ge \cdots \ge \sigma_k(\mb P) \ge 0$, where $\sigma_i(\mb P) $ is the $ith$ largest singular value of $P$.  Clearly, $\sigma_1(\mb P)=1$. From the expression \eqref{eqn:ill}, we can see that the speed of convergence depends on how fast $\mb P^t$ converges to $\frac{1}{k} \bm 1 \bm 1^\top,$ and based on the Perron-Frobenius theory, we have,
\[
\norm{\mb P^t \bm x - \frac{1}{n} \bm 1}{2} \le \sigma_2(\mb P)^t,
\]
for any $\bm x$ in the $k$-dimensional probability simplex. Therefore, the network with large spectral gap, $1-\sigma_2(\mb P)$, is greatly desired. For additional discussions on the importance of this spectral gap, please refer to \cite{duchi2012dual}.

We calculate the spectral gap as $1 - \sigma_2(\mb P)$, where $\sigma_2(\mb P)$ is the second largest singular value of $P$. The $P$ matrix is defined as $A/d$, where $A$ is the adjacency matrix (including self-loop) and $d$ is the in-degree (including self-loop). The spectral gap here is defined as $\sigma_1(\mb P) - \sigma_2(\mb P)$. But $\sigma_1(\mb P)$ the largest singular value should be $1$. So the gap equals $1-\sigma_2(\mb P)$, where $\sigma_2(\mb P)$ is the second largest singular value of $P$.

Hence, \reducename\ network communication graphs with large spectral gap values will converge rapidly. Hence, one should construct sparse network communication topologies for reduce such that the communication costs are low while ensuring large possible spectral gap values of the network.

\end{appendices}

\small
\bibliography{references}

\begin{thebibliography}{10}

\bibitem{apachehama:web}
{{Apache Hama: Framework for Big Data Analytics}}.
\newblock \url{http://hama.apache.org/}.

\bibitem{gpudirect:web}
{{NVIDIA GPUDirect}}.
\newblock \url{https://developer.nvidia. com/gpudirect}.

\bibitem{pascal:web}
{{PASCAL Large Scale Learning Challenge}}.
\newblock \url{http://largescale.ml.tu-berlin.de}, 2009.

\bibitem{cifar10:web}
{{The CIFAR-10 dataset}}.
\newblock \url{https://www.cs. toronto.edu/~kriz/cifar.html}, 2009.

\bibitem{abaditensorflow}
M.~Abadi and et. al.
\newblock Tensorflow: Large-scale machine learning on heterogeneous systems,
  2015.
\newblock {\em Software available from tensorflow. org}.

\bibitem{agarwal2014reliable}
A.~Agarwal, O.~Chapelle, M.~Dud{\'\i}k, and J.~Langford.
\newblock A reliable effective terascale linear learning system.
\newblock {\em JMLR}, 2014.

\bibitem{batson2013spectral}
J.~Batson, D.~A. Spielman, N.~Srivastava, and S.-H. Teng.
\newblock Spectral sparsification of graphs: theory and algorithms.
\newblock {\em Communications of the ACM}, 56(8):87--94, 2013.

\bibitem{bottou-2010}
L.~Bottou.
\newblock Large-scale machine learning with stochastic gradient descent.
\newblock In {\em Springer COMPSTAT}, pages 177--187, Paris, France, 2010.

\bibitem{canny:butterfly}
J.~Canny and H.~Zhao.
\newblock Butterfly mixing: Accelerating incremental-update algorithms on
  clusters.
\newblock In {\em SDM}, pages 785--793, 2013.

\bibitem{chaiken2008scope}
R.~Chaiken, B.~Jenkins, P.-{\AA}. Larson, B.~Ramsey, D.~Shakib, S.~Weaver, and
  J.~Zhou.
\newblock Scope: easy and efficient parallel processing of massive data sets.
\newblock {\em Proceedings of the VLDB Endowment}, 1(2):1265--1276, 2008.

\bibitem{chen2016revisiting}
J.~Chen, R.~Monga, S.~Bengio, and R.~Jozefowicz.
\newblock Revisiting distributed synchronous sgd.
\newblock {\em ICLR Workshop}, 2016.

\bibitem{chen2015mxnet}
T.~Chen, M.~Li, Y.~Li, M.~Lin, N.~Wang, M.~Wang, T.~Xiao, B.~Xu, C.~Zhang, and
  Z.~Zhang.
\newblock Mxnet: A flexible and efficient machine learning library for
  heterogeneous distributed systems.
\newblock {\em arXiv preprint arXiv:1512.01274}, 2015.

\bibitem{chilimbi2014adam}
T.~Chilimbi, Y.~Suzue, J.~Apacible, and K.~Kalyanaraman.
\newblock {Project Adam: Building an Efficient and Scalable Deep Learning
  Training System}.
\newblock In {\em USENIX OSDI}, 2014.

\bibitem{chowdhury2014efficient}
M.~Chowdhury, Y.~Zhong, and I.~Stoica.
\newblock Efficient coflow scheduling with varys.
\newblock In {\em ACM SIGCOMM Computer Communication Review}. ACM, 2014.

\bibitem{chu2007map}
C.~Chu, S.~K. Kim, Y.-A. Lin, Y.~Yu, G.~Bradski, A.~Y. Ng, and K.~Olukotun.
\newblock Map-reduce for machine learning on multicore.
\newblock {\em NIPS}, 19:281, 2007.

\bibitem{chu2003optimizing}
L.~Chu, H.~Tang, T.~Yang, and K.~Shen.
\newblock Optimizing data aggregation for cluster-based internet services.
\newblock In {\em ACM SIGPLAN Symposium on Principles and practice of parallel
  programming {(PPoPP)} , 2003.} ACM.

\bibitem{cipar2013solving}
J.~Cipar, Q.~Ho, J.~K. Kim, S.~Lee, G.~R. Ganger, G.~Gibson, K.~Keeton, and
  E.~Xing.
\newblock Solving the straggler problem with bounded staleness.
\newblock In {\em USENIX HotOS}, 2013.

\bibitem{collobert2011torch7}
R.~Collobert, K.~Kavukcuoglu, and C.~Farabet.
\newblock Torch7: A matlab-like environment for machine learning.
\newblock In {\em BigLearn, NIPS Workshop}, 2011.

\bibitem{condie2010mapreduce}
T.~Condie, N.~Conway, P.~Alvaro, J.~M. Hellerstein, K.~Elmeleegy, and R.~Sears.
\newblock Mapreduce online.
\newblock In {\em NSDI}, volume~10, page~20, 2010.

\bibitem{cui2014exploiting}
H.~Cui, J.~Cipar, Q.~Ho, J.~K. Kim, S.~Lee, A.~Kumar, J.~Wei, W.~Dai, G.~R.
  Ganger, P.~B. Gibbons, et~al.
\newblock Exploiting bounded staleness to speed up big data analytics.
\newblock In {\em USENIX ATC}, 2014.

\bibitem{cui2016geeps}
H.~Cui, H.~Zhang, G.~Ganger, P.~Gibbons, and E.~Xing.
\newblock Geeps: Scalable deep learning on distributed gpus with a
  gpu-specialized parameter server.
\newblock In {\em Proceedings of the Eleventh European Conference on Computer
  Systems}. ACM, 2016.

\bibitem{dai2013petuum}
W.~Dai, J.~Wei, X.~Zheng, J.~K. Kim, S.~Lee, J.~Yin, Q.~Ho, and E.~P. Xing.
\newblock Petuum: A framework for iterative-convergent distributed ml.
\newblock {\em arXiv preprint arXiv:1312.7651}, 2013.

\bibitem{dean2012large}
J.~Dean, G.~Corrado, R.~Monga, K.~Chen, M.~Devin, Q.~V. Le, M.~Z. Mao,
  M.~Ranzato, A.~W. Senior, P.~A. Tucker, et~al.
\newblock Large scale distributed deep networks.
\newblock In {\em NIPS}, 2012.

\bibitem{dean2008mapreduce}
J.~Dean and S.~Ghemawat.
\newblock Mapreduce: simplified data processing on large clusters.
\newblock {\em Communications of the ACM}, 51(1):107--113, 2008.

\bibitem{duchi2012dual}
J.~C. Duchi, A.~Agarwal, and M.~J. Wainwright.
\newblock Dual averaging for distributed optimization: convergence analysis and
  network scaling.
\newblock {\em Automatic control, IEEE Transactions on}, 57(3):592--606, 2012.

\bibitem{gabriel2004open}
E.~Gabriel, G.~E. Fagg, G.~Bosilca, T.~Angskun, J.~J. Dongarra, J.~M. Squyres,
  V.~Sahay, P.~Kambadur, B.~Barrett, A.~Lumsdaine, et~al.
\newblock {Open MPI: Goals, concept, and design of a next generation MPI
  implementation}.
\newblock In {\em Recent Advances in Parallel Virtual Machine and Message
  Passing Interface}, pages 97--104. Springer, 2004.

\bibitem{gemulla2011large}
R.~Gemulla, E.~Nijkamp, P.~J. Haas, and Y.~Sismanis.
\newblock Large-scale matrix factorization with distributed stochastic gradient
  descent.
\newblock In {\em ACM KDD}, pages 69--77, 2011.

\bibitem{ghoting2011systemml}
A.~Ghoting, R.~Krishnamurthy, E.~Pednault, B.~Reinwald, V.~Sindhwani,
  S.~Tatikonda, Y.~Tian, and S.~Vaithyanathan.
\newblock Systemml: Declarative machine learning on mapreduce.
\newblock In {\em Data Engineering (ICDE), 2011 IEEE 27th International
  Conference on}, pages 231--242. IEEE, 2011.

\bibitem{goiri2015approxhadoop}
{\'I}.~Goiri, R.~Bianchini, S.~Nagarakatte, and T.~D. Nguyen.
\newblock Approxhadoop: Bringing approximations to mapreduce frameworks.
\newblock In {\em Proceedings of the Twentieth International Conference on
  Architectural Support for Programming Languages and Operating Systems}, pages
  383--397. ACM, 2015.

\bibitem{gonzalez2015asynchronous}
J.~E. Gonzalez, P.~Bailis, M.~I. Jordan, M.~J. Franklin, J.~M. Hellerstein,
  A.~Ghodsi, and I.~Stoica.
\newblock Asynchronous complex analytics in a distributed dataflow
  architecture.
\newblock {\em arXiv preprint arXiv:1510.07092}, 2015.

\bibitem{gonzalez2012powergraph}
J.~E. Gonzalez, Y.~Low, H.~Gu, D.~Bickson, and C.~Guestrin.
\newblock Powergraph: Distributed graph-parallel computation on natural graphs.
\newblock In {\em USENIX OSDI}, 2012.

\bibitem{gregor2005parallel}
D.~Gregor and A.~Lumsdaine.
\newblock {The parallel BGL: A generic library for distributed graph
  computations}.
\newblock {\em Parallel Object-Oriented Scientific Computing (POOSC)}, 2:1--18,
  2005.

\bibitem{guo2012spotting}
Z.~Guo, X.~Fan, R.~Chen, J.~Zhang, H.~Zhou, S.~McDirmid, C.~Liu, W.~Lin,
  J.~Zhou, and L.~Zhou.
\newblock Spotting code optimizations in data-parallel pipelines through
  periscope.
\newblock In {\em USENIX OSDI}, 2012.

\bibitem{isard2007dryad}
M.~Isard, M.~Budiu, Y.~Yu, A.~Birrell, and D.~Fetterly.
\newblock Dryad: distributed data-parallel programs from sequential building
  blocks.
\newblock In {\em ACM EuroSys}, 2007.

\bibitem{kalia2014using}
A.~Kalia, M.~Kaminsky, and D.~G. Andersen.
\newblock {Using RDMA efficiently for key-value services}.
\newblock In {\em SIGCOMM}, pages 295--306. ACM, 2014.

\bibitem{kumar2016hold}
G.~Kumar, G.~Ananthanarayanan, S.~Ratnasamy, and I.~Stoica.
\newblock Hold’em or fold’em? aggregation queries under performance
  variations.
\newblock 2016.

\bibitem{li2015malt}
H.~Li, A.~Kadav, E.~Kruus, and C.~Ungureanu.
\newblock Malt: distributed data-parallelism for existing ml applications.
\newblock In {\em Eurosys}. ACM, 2015.

\bibitem{li2014parameterserver}
M.~Li, D.~Andersen, A.~Smola, J.~Park, A.~Ahmed, V.~Josifovski, J.~Long,
  E.~Shekita, and B.-Y. Su.
\newblock Scaling distributed machine learning with the parameter server.
\newblock In {\em USENIX OSDI}, 2014.

\bibitem{li2013distributed}
M.~Li, D.~G. Andersen, and A.~Smola.
\newblock Distributed delayed proximal gradient methods.
\newblock In {\em NIPS Workshop on Optimization for Machine Learning}, 2013.

\bibitem{malewicz2010pregel}
G.~Malewicz, M.~H. Austern, A.~J. Bik, J.~C. Dehnert, I.~Horn, N.~Leiser, and
  G.~Czajkowski.
\newblock Pregel: a system for large-scale graph processing.
\newblock In {\em Proceedings of the 2010 ACM SIGMOD International Conference
  on Management of data}, pages 135--146. ACM, 2010.

\bibitem{mcsherry2015scalability}
F.~McSherry, M.~Isard, and D.~G. Murray.
\newblock Scalability! but at what cost?
\newblock In {\em 15th Workshop on Hot Topics in Operating Systems (HotOS XV)},
  2015.

\bibitem{frank:graphprocessing}
{McSherry, Frank}.
\newblock {{Progress in graph processing: Synchronous vs asynchronous graph
  processing}}.
\newblock \url{https://github.com/frankmcsherry/blog
  /blob/master/posts/2015-12-24.md}.

\bibitem{murray2013naiad}
D.~G. Murray, F.~McSherry, R.~Isaacs, M.~Isard, P.~Barham, and M.~Abadi.
\newblock Naiad: A timely dataflow system.
\newblock In {\em ACM SOSP}, 2013.

\bibitem{milde:web}
{NEC Laboratories America}.
\newblock {{MiLDE: Machine Learning Development Environment}}.
\newblock \url{http://www.nec-labs.com/research-departments
  /machine-learning/machine-learning-software/ Milde}.

\bibitem{nedic2009distributed}
A.~Nedic and A.~Ozdaglar.
\newblock Distributed subgradient methods for multi-agent optimization.
\newblock {\em IEEE Transactions on Automatic Control}, pages 48--61, 2009.

\bibitem{nightingale2012flat}
E.~B. Nightingale, J.~Elson, J.~Fan, O.~Hofmann, J.~Howell, and Y.~Suzue.
\newblock Flat datacenter storage.
\newblock In {\em Presented as part of the 10th USENIX Symposium on Operating
  Systems Design and Implementation (OSDI 12)}, pages 1--15, 2012.

\bibitem{noel2014dogwild}
C.~Noel and S.~Osindero.
\newblock Dogwild!—distributed hogwild for cpu \& gpu.
\newblock In {\em NIPS workshop on Distributed Machine Learning and Matrix
  Computations}, 2014.

\bibitem{olston2008pig}
C.~Olston, B.~Reed, U.~Srivastava, R.~Kumar, and A.~Tomkins.
\newblock Pig latin: a not-so-foreign language for data processing.
\newblock In {\em Proceedings of the 2008 ACM SIGMOD international conference
  on Management of data}, pages 1099--1110. ACM, 2008.

\bibitem{ousterhout2015making}
K.~Ousterhout, R.~Rasti, S.~Ratnasamy, S.~Shenker, B.-G. Chun, and V.~ICSI.
\newblock Making sense of performance in data analytics frameworks.
\newblock In {\em NSDI}, 2015.

\bibitem{ouyang2013stochastic}
H.~Ouyang, N.~He, L.~Tran, and A.~Gray.
\newblock Stochastic alternating direction method of multipliers.
\newblock In {\em Proceedings of the 30th International Conference on Machine
  Learning}, pages 80--88, 2013.

\bibitem{power2010piccolo}
R.~Power and J.~Li.
\newblock Piccolo: Building fast, distributed programs with partitioned tables.
\newblock In {\em USENIX OSDI}, pages 293--306, 2010.

\bibitem{recht2011hogwild}
B.~Recht, C.~Re, S.~Wright, and F.~Niu.
\newblock {Hogwild: A lock-free approach to parallelizing stochastic gradient
  descent}.
\newblock In {\em NIPS}, 2011.

\bibitem{shatdal1995adaptive}
A.~Shatdal and J.~F. Naughton.
\newblock Adaptive parallel aggregation algorithms.
\newblock In {\em ACM SIGMOD Record}, volume~24, pages 104--114. ACM, 1995.

\bibitem{simonyan2014very}
K.~Simonyan and A.~Zisserman.
\newblock Very deep convolutional networks for large-scale image recognition.
\newblock {\em arXiv preprint arXiv:1409.1556}, 2014.

\bibitem{tran2009sybil}
D.~N. Tran, B.~Min, J.~Li, and L.~Subramanian.
\newblock Sybil-resilient online content voting.
\newblock In {\em NSDI}, volume~9, pages 15--28, 2009.

\bibitem{valadarsky2015xpander}
A.~Valadarsky, M.~Dinitz, and M.~Schapira.
\newblock Xpander: Unveiling the secrets of high-performance datacenters.
\newblock In {\em Proceedings of the 14th ACM Workshop on Hot Topics in
  Networks}, page~16. ACM, 2015.

\bibitem{valiant1990bridging}
L.~G. Valiant.
\newblock A bridging model for parallel computation.
\newblock {\em Communications of the ACM}, 33(8):103--111, 1990.

\bibitem{verma2013breaking}
A.~Verma, B.~Cho, N.~Zea, I.~Gupta, and R.~H. Campbell.
\newblock Breaking the mapreduce stage barrier.
\newblock {\em Cluster computing}, 16(1):191--206, 2013.

\bibitem{wang2013asynchronous}
G.~Wang, W.~Xie, A.~J. Demers, and J.~Gehrke.
\newblock Asynchronous large-scale graph processing made easy.
\newblock In {\em CIDR}, 2013.

\bibitem{yu2008dryadlinq}
Y.~Yu, M.~Isard, D.~Fetterly, M.~Budiu, {\'U}.~Erlingsson, P.~K. Gunda, and
  J.~Currey.
\newblock Dryadlinq: A system for general-purpose distributed data-parallel
  computing using a high-level language.
\newblock In {\em USENIX OSDI}, 2008.

\bibitem{yun2014nomad}
H.~Yun, H.-F. Yu, C.-J. Hsieh, S.~Vishwanathan, and I.~Dhillon.
\newblock {NOMAD: Non-locking, stOchastic Multi-machine algorithm for
  Asynchronous and Decentralized matrix completion}.
\newblock In {\em ACM VLDB}, 2014.

\bibitem{zaharia2012resilient}
M.~Zaharia, M.~Chowdhury, T.~Das, A.~Dave, J.~Ma, M.~McCauley, M.~J. Franklin,
  S.~Shenker, and I.~Stoica.
\newblock Resilient distributed datasets: A fault-tolerant abstraction for
  in-memory cluster computing.
\newblock In {\em USENIX NSDI}, 2012.

\bibitem{zhang2015poseidon}
H.~Zhang, Z.~Hu, J.~Wei, P.~Xie, G.~Kim, Q.~Ho, and E.~Xing.
\newblock Poseidon: A system architecture for efficient gpu-based deep learning
  on multiple machines.
\newblock {\em arXiv preprint arXiv:1512.06216}, 2015.

\end{thebibliography}
\bibliographystyle{abbrv} 

\end{document}